\documentstyle[11pt,epsf]{mnthesis}
\newcommand{\bb}{\begin{equation}}   
\newcommand{\ee}{\end{equation}}

\begin{document}
\bibliographystyle{plain}
%\include epsf
%  Title and other sections that come before the body  of the document
%     \include{title}
%     \newpage
%     \pagenumbering{roman}
%     \include{introduction}
%     \tableofcontents
%     \listoffigures
%     \listoftables
% 
%   Craig's figure inclusion stuff 
%   note put in parameters 1-4 with {}
%
%\def\myfig#1#2#3#4{
%    \vspace{.05in}
%    \begin{figure}
%  
%    \hspace{#4in}\special{include #1}
%    \vspace{#3in} 
%    \vspace{.05in}
%    \caption{#2}
%\end{figure}
%}
%\def\nofig#1#2#3#4{
%    \vspace{.05in}    
%    \begin{figure}
%    \vspace{#3in}
%    \hspace{#4in}
%    \vspace{.05in}
%    \caption{#2}
%\end{figure}
%}
% Now lets include the body of the document...
%  \newpage
%  \pagenumbering{arabic}
%     \include{"chapter1"}
%     \include{chapter2}
%     \include{chapter3}
%     \include{chapter4}
%     \include{chapter5}
%     \include{chapter6}
%     \include{chapter7}
%     \include{chapter8} 
%     \include{chapter9}
%     \include{chapter10}
%     \include{summary}
%     \include{bibliography}     
% In the appendix some things should change, so a user should say that
% its starting.
%     \appendix
%     \include{appa}
%\end{document}

% For a  PhD give the command \phd. If you don't do this then the default
% is for a Masters Degree.
% Optional fields are (where normal is for PhD or the default Master
% of Science:
%	\degree (normally Doctor of Philosophy or Master of Science)
%	\initials (normally Ph.D. or M.S.)
%	\draft (this should probably be set  until near the final
%              drafts. Currently it modifies the title page.)
    \phd
%   \draft
    \title{\bf The Semileptonic Lifetime of the D Meson\\}
    \author{Richard David Dikeman}
    \campus{Theoretical Physics Institute} 
    \program{Physics} 
    \director{Mikhail Shifman}
%   \words{116}
    \approvals{4}
%   \copyrightpage % Do you want copyright protection?
    \submissionmonth{February} 
    \submissionyear{1998}
    \abstract{The full treatment, both nonperturbative - through $O(1/m_c^3)$ -
and perturbative - through $O(\alpha_s)$ and BLM type 
$O(\alpha_s^2)$ - of 
the semileptonic lifetime of the D meson is given.  A
dedicated discussion of the numerics involved is made,
with the result that the
leading order prediction plus corrections fail to match experiment. To 
explain
the difference, a model of duality is invoked, where it is shown that the
asymptotic nature of the nonperturbative expansion may be responsible for the
theoretical experimental discrepancy. A particularly interesting 
conclusion which follows from the model is that any attempts to extract 
$\alpha_s$ from 
$\tau$ decays are doomed, since the same duality violating mechanism is 
at work here.

}
    \introduction{\begin{center}
``All is mystery; but he is a slave who will not struggle to penetrate
the dark veil'' - Disraeli
\end{center}

Hopefully, there will be a few people outside the high energy theory  
communitity who will at least take a glance at this thesis. This 
introduction is written for them, in hope that all the complicated equations,
diagrams, and verbage following will have some general meaning. I think it was
Feynman who said that you should be able to explain your research to a seventh 
grader, or something like that. How to DO the research, is, of course, another
matter all together, but what the problem is, and what its solution is should 
be understandable to anyone. This is my attempt to explain:

Most people know that there are things called atoms, and have seen 
simple diagrams (resembling the solar system to some extent) 
which picture electrons whirling around a nucleus.
Although, after a semester 
of quantum theory, one would most likely cringe at such a diagram, these 
simplified pictures do in fact illustrate some key features of the 
atom. The electrons are usually displayed either as 
clouds, illustrating the wave nature of the very small, or as particles 
traversing orbits around a center, the nucleus. Usually, the electron, 
when pictured in the `planet' like orbital picture, is shown as a `dot'. 
In some sense, this is a fair representation. As far as we know at present, 
the electron is a particle which has no structure - no radius whatsoever, and 
no constitutents. The electron is, as far as we can tell, the perfect 
physical representation of a mathematical point.
The nucleus, however, is an entirely different story. 
It has constituents, being made up of protons and neutrons. Surprisingly, 
although 
protons are positive in charge (neutrons are neutral!) the nucleus does 
not `fly' apart (like charges repel). There is another force binding 
the nucleus together, stronger than the electrostatic repulsion, 
called the strong force. The modern day 
theory of the strong force is called QCD - Quantum Chromodynamics, which 
is the topic of my thesis. The `objects' of QCD, however, are not the 
proton and neutron. It is known 
for some decades now that the proton and neutron are not at all like the 
electron. These particles {\it{do}} have internal structure. Indeed, the 
proton and neutron 
(and many other particles like them called either mesons or baryons) are 
made 
from particles called quarks and gluons - the true `objects' of QCD. 
Quarks and gluons themselves are like the electron - 
seemingly pointlike. Quarks come in different `flavors' (i.e. there are 
different kinds which differ by their masses). So far, there are six 
different flavors found: up, down, strange, charm, bottom, and top. 
Quarks are never found alone, but are confined either in pairs 
(mesons) or 
triplets (baryons). As stated above, there is a theory - QCD - which 
mathematically details the 
dynamics of how quarks interact. What does it mean to say that there is a 
`theory' detailing the dynamics? It basically means we should be able to 
calculate some property of quarks, and compare it to experiment. For 
example, this thesis is concerned with calculating the lifetime (the 
strange, charmed, bottom, and top quark are all unstable, and quickly 
decay when created at a particle accelerator) of a a meson called the D 
meson (the D meson is a pair of quarks - one charmed and one up or down).
Specifically, I calculate the decay of the D meson into a pair of leptons 
(an electron is a member of the lepton `family')
and all possible hadrons (a hadron is either a 
meson or a baryon, but in general I mean here all other possible 
combinations of quarks).
The D meson lifetime is well established experimentally. Calculating it, 
on the other hand, is absolutely difficult. QCD is a notoriously difficult 
theory to calculate with. Noble Prize winner Steven Weinberg is quoted as 
saying: 

\begin{center}
``There's a long tradition in theoretical physics, which by no 
means affected everyone but certainly affected me, that said the strong 
interactions are too complicated for the human mind.''
\end{center}

Indeed, making an 
exact calculation of the decay time of the D meson, or 
any other meson will probably {\it{never}} happen. This is really not 
such a big 
deal, however, as this is almost always the 
case in physics anyway. Hardly anything can be calculated without making 
some approximations, and when it can, it usually doesn't correspond to the 
true physical world anyway. (Of course, nature doesn't depend on 
whether we can `calculate' it or not - e.g. atoms do not calculate 
their own behavior!, still much of the richness of our universe resides 
in the fact that nature is calculable only in approximation. If nature 
happened to correspond to  a very simple theory which could be 
used to calculate for {\it{all}} physical phenomenon, we would live in a 
boring place indeed. (Let me remark that it 
seems certain that nature abides by some very elegant, and simple 
theories which lead to, most of the time, very rich phenomena. It is 
usually very `easy' to write the theory, and most often very difficult to 
calculate with it - thus the need for approximations!)) 
The D meson is 
seemingly an ideal situation for 
making approximations. The reason is that the charmed quark has a high 
mass. What can be done then computationally, is to express the decay 
lifetime in a power series:
\bb
\Gamma = a_0 + a_1 (\frac{\mu}{m_c}) + a_2 (\frac{\mu}{m_c})^2 + 
a_3 (\frac{\mu}{m_c})^3 ...
\ee
here $m_c$ is the mass of the charm quark, and $\mu$ is a tool of the 
theorist, and is chosen so that the expansion 
converges (there are subtleties here which even real experts do not get!). 
Numerically, $\mu$ is chosen so that $\mu/m_c = .5$ or so. 
Suppose then, that $a_0, a_1, a_2,...$ are - all different - but 
all close to each other (unless there is some strong reason, this could be 
expected - a statement of vague intuition to be sure). Then, for each new 
term in the series, say $a_4$, $a_5$, etc. the new term is also multiplied 
by increasing powers of $\mu/m_c=.5$. 
Technically, it is impossible to calculate the full series - no one knows 
how, but we can approximate the result by calculating 
the first few terms and then truncating the series when each new term gets 
insignificant (or too hard!). 
The calculation of $a_0$, $a_1$, and $a_2$ were all completed before I 
knew what a heavy quark even was, but not much before! Check out the 
result of this calculation - equation 57, and in particular note that $m_c$ 
is in the 
denominator of the last few terms, in tune with what is written above.
Essentially, the first five chapters (aside from the 
preface) are concerned with this computation. All the details of `how to 
do it' are here. Chapter 6 and 7 explain the actual numerical values that 
are 
used to check and see if equation 5.22 actually matches experiment. This is 
an interesting area of scientific debate. As it turns out, if one measures
numerical values for the charmed quark mass and other parameters from 
experiments and applies them to the problem of the D lifetime, the 
theoretical value of the decay time is roughly 50 percent of 
the experimental value (pretty bad agreement) - check out equation 7.3. 
The key numerical parameter here is $m_c$, since, as one can see from 
equation 5.22, it enters into the expression of the lifetime as $m_c^5$.
Because $m_c$ is to the fifth power, only a slight increase, roughly 10 
percent, of  it's value will compensate the 50 percent deficit between 
theory and experiment.
However, those (Dikeman included) who think 
that $m_c$ should have a {\it{low}} value and thus spoil the agreement
between theory and experiment regarding the D lifetime 
believe this because it would spoil too many other results which say $m_c$ 
is low. Thus, there is a problem.

One possible next move is to try and calculate the next 
term in the series and see if it somehow can raise the decay value a 
whole 50 percent - an unlikely scenario since remember, each new term 
should be .5 times smaller than its previous one, and thus numerically - 
check out equation 88 - we would expect, at best, this `cubic' term in the 
series to contribute at most 10 percent, but not 50 percent.
Still, though, one must try, and so 
the calculation of the next term in the series, $a_3$, was 
my first task as a graduate student. It was 
hoped that this `cubic' 
term would explain the puzzle. This is the discussion of chapter 8.
Well, after a lot of work, no 
such luck. 
My advisor Misha Shifman, his past student Boris Blok, and myself wrote a 
paper demonstrating that the next term in the series doesn't help 
matters, and may even make them worse. At least in this paper we tried to 
give a number of possible 
explanations why the puzzle existed - our main guess was that the charmed 
quark was not quite heavy enough for the power series expansion to work 
in the first place. Despite the negative results, though, the 
paper was mildly successful - currently it has been cited in other papers 
17 times, 
which is not at all bad. 

With the puzzle deepened, I went on to another 
project, but then returned to the D meson puzzle a year later. My 
advisor, Misha Shifman, 
suggested to me and fellow graduate student and friend Boris Chibisov while 
in a summer school at 
Boulder, Colorado that there may be a way of estimating what is 
`leftover' when truncating the series, i.e. estimating the remaining 
pieces, $a_5$ ... $a_{10}$, etc. of the 
series. This work was the first real attempt at such an estimate 
concerning the D meson, or any other QCD phenomenon, and so is 
a much more general paper than the first one, and a bit more successful 
- it 
has 21 citations so far, and I expect a lot more in the future. The ideas 
of this paper are presented in chapter 10. 
In general, these ideas are interesting, but they don't exactly solve the 
remaining puzzle, instead putting it in proper perspective. As I allude to 
above, it is difficult to solve a hard problem without making 
approximations, and chapter 10 is loaded with them, but we at least hope 
to correctly capture some important features of the real, if there ever 
really is one!, solution to the puzzle. In addition, the model developed 
in chapter 10 has 
strong implications for other QCD processes as well, and sheds a lot of 
doubt on a few particular phenomenological studies which are quite important.
And thats the end of the story.

}
    \dedication{\begin{center}
``Born under widespread changes to search for...higher reason. Learning 
the ropes ok, but fate just runs you around.'' Jay Farrar of {\it{Son Volt}}
\end{center}

So many people have been instrumental in my attaining this goal that 
I shudder to write the usual, brief: e.g. `To my parents'. This is a 
great chance to thank family, and friends, and I want to use it and do a 
proper job. 

Minnesota is an unspeakably cold, dark place, but the 
opportunities to have a great time still abound. At my side, participating 
in most of these great times over the last five years has been John 
Capriotti 
and so first and foremost, I'd like to thank him. 
The great music, suds, untold number of laughs, 
stratocaster jam sessions, and Slim Dunlap shows,
all while sorting out the meaning of 
life, shared with Cap over the last few years have turned what would have 
been just a remarkable experience in the world of physics, into a rich life 
experience.  

Mostly responsible for my remarkable physics experience noted 
above is Mikhail Shifman. When I came to Minnesota from Ann Arbor, I laid down 
to go to sleep on one of my first nights and thought ``five years or so from 
now, I'll be a Ph.D. in physics!'' From my first days of discovering 
science by myself in popular books by Isaac Asimov, Carl Sagan, etc., I 
have always loved science, and so the 
thought was formed not just from the desire to obtain a Ph.D. but to 
{\it{really}} learn about the universe. After a dissapointing first year in 
which no experimental condensed matter theorist would take me (!),  
Misha, after my successful tour of duty in his Quantum Mechanics class, 
gave me some problem involving heavy quarks (uhh... what?!). 
Now, I have some 7 or so papers in the field of QCD, and have to laugh at 
that memory (I really had no idea just who Misha was: `Sum Rules? 
What are those?' and what kind of education I had in store for me). The 
thrill of working with Misha comes from learning what it means to have 
physical insight into nature. Because of this I was afforded to work on some 
really neat problems and learn a helluva lot more about science than I 
had ever dreamed. To return in time to that guy (me!), about to go to 
sleep and excited about the coming five years 
and tell him that I'd be working with one of the top theorists in 
particle physics, eventually traveling the world giving seminars here and 
there, 
and producing important work....WOW! - It's really been an extraordinary 
experience. Thanks Misha. 

I also want to thank the other guys I worked with during my physics 
career, especially Kolya Uraltsev. I only wish I had half as much passion 
for physics as Kolya, even though after all our discussions, I found that 
my own had at least doubled! Kolya was one of those guys so necessary in 
being able to do any reasonable work if you have modest ability - a guy 
you can ask absolutely dumb questions to! Thanks Kolya. Also, fellow grad 
students Boris Chibisov, Jamal Marian, 
Alejandro Mercado, Dick Madden, Bayram Tekin, Lev Koyrakh, Denne 
Weslowski, Igor Zutic, Joel Kindem, Jason Sielaff, and Emil Akhmedov were 
great resources and fun.

I'd also like to thank my high school 
and college friends: Huey, Voll, 
Fritz, Bones, Nip, Shankar, Chip, Fuzz, and Klaz. With friends like these, I 
can confidently say that few have ever had as many laughs and good times 
as I have over the last dozen years.

Lastly, I'd like to thank my parents and sister  
who instilled the self confidence and ability in me to take on such 
a task and complete it successfully. 
More than anything else, I'm lucky to be in such a happy, loving family.

}
    \figurespage
%   \tablespage
%
% The \beforepreface  command actually causes insertion of the title, 
% abstract,  and copyright pages into the new document.

\beforepreface 

%
% The \afterpreface  command actually causes insertion of the
% contents, list of figures, etc. into the new document.
\afterpreface

     \newpage
     \pagenumbering{roman}

  \def\myfig#1#2#3#4{
    \vspace{.05in}
    \begin{figure}

    \hspace{#4in}\special{include #1}
    \vspace{#3in}
    \vspace{.05in}
    \caption{#2}
\end{figure}
}
\def\nofig#1#2#3#4{
    \vspace{.05in}
    \begin{figure}
    \vspace{#3in}
    \hspace{#4in}
    \vspace{.05in}
    \caption{#2}
\end{figure}
}  
 \newpage
  \pagenumbering{arabic}

\chapter{Preface} \begin{center} ``The age in which we live is the age in
which we are discovering the fundamental laws of nature, and that day will
never come again.'' - Feynman \end{center}

\begin{center}
``The more the universe seems comprehensible, the more it also seems 
pointless. But if there is no solace in the fruits of our research, there 
is at least some consolation in the research itself. Men and women are 
not content to comfort themselves with tales of gods and giants, or to 
confine their thoughts to the daily affairs of life; they also build 
telescopes and satellites and accelerators, and sit at their desks for 
endless hours working out the meaning of the data they gather. The effort 
to understand the universe is one of the very few things that lifts human 
life a little above the level of farce, and gives it some of the grace of 
tragedy'' - Steven Weinberg
\end{center}

In any physical theory, it is often helpful, sometimes crucial and 
perhaps even mandatory, to identify the simplest problem the theory can 
solve.
Fermi, referring to this feature of theoretical physics, liked to 
pose the question
`What is the hydrogen atom of the theory?'  
For Quantum Chromodynamics (QCD), the theory of nuclear force, this is not 
such an easy question to answer. 
Historically, as a shockingly rich spectra of 
hadrons was discovered by experimentalists in the 
50's, theorists attacked problems of strong resonance physics not with 
powerful analytic 
results from first principles QCD - QCD did not even exist until 1973, but 
instead with ideas based on symmetries. The group (flavor) structure 
of QCD was revealed by the spectroscopy of 
hadrons. This is the work most associated with Gell-Mann and 
Ne'eman \cite{8fold}.
Gell-Mann realized that the spectra of hadrons - protons, neutrons, and the 
myriad resonances being discovered - could be explained by introducing 
\cite{GMtalk} 
what he deemed fictitious entities: 
quarks, which carried a color charge, and some `glue' to hold the quarks 
together.  
Aside from Gell-Mann's `Eight-fold way' which, anyway, dealt with the 
flavor chiral symmetry \cite{WEINBE} of 
QCD, and not its more fundamental color symmetry,
the answer to Fermi's question might be, at least regarding the gauge 
structure of the theory,
the special puzzle 
of the $\Delta^{++}$ wave function. 

The $\Delta^{++}$ is a spin 3/2 particle, a fermion, but with entirely 
symmetric wave-function. 
As is well known, a half-integer spin particle necessarily has an 
anti-symmetric wave function, and so there is a puzzle.
Instead of appealing to some perverse solution, the elegant notion of a 
color charge, yielding an 
antisymmetric, colorless wave function was put forth which 
solved this and other such problems \cite{color}.
Thus the idea that quarks carried color, and appeared in nature in only 
confined combinations 
allowing a color-singlet wavefunction was born, and so was QCD 
\cite{GMtalk}, \cite{GWP}. In this sense, i.e.  elucidating 
that there is a color wave function, and thus an $SU(3)$ non-Abelian gauge 
theory,
\cite{YM}
the $\Delta^{++}$ {\it{is}} the hydrogen atom of QCD. But when Fermi asked 
this question, surely he 
was referring to a more dynamical problem in the theory - the 
$\Delta^{++}$ has helped us 
{\it{identify}} the theory, but now we should {\it{calculate}} the 
spectrum of the simplest system with it.

Possibly the most genuinally interesting thing 
about QCD is that there are such a variety of dynamical 
situations where we in fact 
have a simple 
hydrogen atom like problem, which quickly becomes contaminated with 
the intricacies 
of QCD. Analyzing phenomena such as $e^+ e^-$ annhilation, DIS, heavy quark 
decays, etc. we, at first,  find quick 
successes with traditional methods only to later suffer severe impedances 
where our methods aren't applicable. At the core of the
simplicities of these systems is asymptotic freedom 
\cite{GWP}.
Asymptotic freedom tells us that we can use our powerful arsenals of 
perturbative methods, 
developed already in the study of QED, to attack problems in the 
ultraviolet (high momentum 
domain) where the QCD coupling constant is small. In the infrared (soft 
momentum domain), 
however, where $\alpha_s \approx 1$ there is what is called infrared 
slavery, and we must resort to other methods. Essentially, asymptotic 
freedom creates a calculational 
boundary, beyond which we must be extremely clever to extract 
model independant results. 
Due to  the nature of the running of the strong coupling constant, we face a 
situation where for soft processes, 
typically at scales around a few GeV, we have no reliable way to 
calculate in QCD. QCD is a nonlinear system with no 
analytic solution, and in the infrared, we have no small quantity that we 
can expand in - a truly  great 
challenge.  

And so, as it turns out, the 
biggest challenge of QCD is to understand its nonperturbative physics. 
Really, then, if we are to 
ask Fermi's question of QCD, cast in the pall of trying to 
understand the genuine solution to the soft problem, we realize that 
perhaps we are stuck! Indeed, 
taking a fairly harsh view of some of the extremely hard, and amazing work
in QCD phenomenology
done in 
the last 20 years on the soft problem, e.g. ref. \cite{SVZ}, one might say 
that the problem of the soft, nonperturbative 
aspects of QCD have,  in some sense, 
merely been swept under the rug! Since the inception of QCD, the primary 
theoretical  tool used to 
deal with the soft physics of QCD has been
Wilson's Operator Product Expansion 
\cite{WOPE}.
Essentially, the central utility of the OPE is that it allows us to
{\it{characterize}} - but not calculate! - the fundamental soft 
quantities of QCD, which we can then extract  
from phenomenology. So we give a name for our ignorances, and can even
ask for them to be measured by experimentalists, but we have no idea 
how to calculate these 
ferocious beasts. With that perspective, perhaps the answer to Fermi's 
question will not 
actually be how to calculate this-or-that process, but instead how to 
calculate this-or-that condensate! Who knows. 

This thesis will not at all be an attempt to answer the question of how to 
calculate the condensates 
in QCD. I, instead, will back down from this Herculean task,  and 
take 
on a lesser trial. 
In this thesis, I will look at the lifetime of the D meson. In 
all hadronic processes 
there is, 
perhaps, no system where one can more easily immerse oneself into the 
intricasies of QCD 
mentioned above.  The D meson, in fact, is much like the hydrogen atom! 
This meson contains a 
heavy quark at its center with a light quark bound to it by an infinite 
number of quark/gluonic 
degrees of freedom. Because the unstable charm quark at the meson's core 
is so heavy, energies 
involved in its decay will be large enough to invoke asymptotic freedom, 
and thus traditional 
perturbative methods. However, since
the charmed quark decays 
in an environment where typical momenta are of the order of 
$\Lambda_{QCD}$ 
(the momenta where the strong coupling constant, $\alpha_s \sim 
1$), we will also have to deal with nonperturbative effects.
By solving the problem of the lifetime of the D meson, hopefully revealing 
some of nature's 
trickeries in the strong interaction section, we also may receive other 
dividends as well - this is 
a weak interaction, and so we might get some information on CP violation 
if we can do the job right.

Not to spoil the story, but as 
it turns out, we won't be able to do the job right. No, in fact, we will learn 
from 
this case that the answer to Fermi's question posed above: `how do we 
calculate this-or-that condensate'  
still might get us nowhere. Indeed, to really get to the bottom of 
QCD we will have to 
know how to calculate the vacuum fluctuations themselves, and how this 
could 
ultimately be done, no one knows. 
Well, actually, it was hoped that instanton physics \cite{BPST} 
might pave the way for a complete understanding here. For instance if we 
knew that the real gauge 
potential in nature was just the one-instanton gauge potential, and we 
knew how to calculate the 
density of states of the instantons, then we could calculate this-or-that 
condensate, on top of 
having direct information about the background field of vacuum 
fluctuations.
By studying the lifetime of the D meson,
we will, at least, learn a few things about the character of these 
fluctuations. The D meson turns 
out to be an interesting system for the following reason: the charm quark 
is both heavy {\it{and}} light, 
so to speak. Heavy because  $\alpha_s(m_c) << 1$ and light since in the 
OPE expansion, $\mu/m_c$ is not small, i.e. $m_c$ is heavy enough for us to 
use 
the OPE expansion, 
$\mu/m_c \sim 0.5$, but it is light enough that this expansion is not so good!
Indeed, at 
the end of the day, we will find that with the constraints imposed on us for 
$m_c$, and other 
pertinent parameters, the standard OPE treatment of the D lifetime FAILS! 
This failure of duality  will 
then be analyzed. In the end, we will find that the D meson is not such a 
good kitchen for making 
the dinner of CP Violation, but {\it{is}} a good kitchen for making the 
breakfast of 
QCD.
So what I will do in this thesis is little more than the standard QCD 
treatment - with a few 
bells and whistles which I helped develop over the past few years - of a 
heavy quark 
decaying with a light spectator.  In the introduction, I will discuss the OPE 
method, 
reviewing how to calculate with it, and explaining why it is so useful for 
QCD, and in particular 
the decay of a 
heavy meson. Then, in the following sections, I will discuss perturbative 
and nonperturbative 
aspects of the problem of the lifetime of the D meson, and all revelant 
features and numerics thereof. In the end, we will be faced with the 
following conclusion: in the semileptonic decay of the D meson, duality 
fails, i.e. theory and experiment simply do not match. This failure 
should be taken as a warning sign for all other hadronic processes below 
$\approx 2 \; {\rm{GeV}}$. For example, we can 
not trust the low energy $\alpha_s$ extractions from $\tau$ decays.

\chapter{Introduction - the OPE}

\begin{center} 
{``So, Dave ... 
you decided to study theoretical particle physics?
Its very, very difficult, you know....'' - Arkady Vainshtein}.
\end {center}

In the treatment of the decay of the D meson, we  rely almost
exclusively on Wilson's 
OPE \cite{WOPE}. 
Wilson's OPE is actually nothing more than a method of calculating 
with a 
quantum field theory. It is just the method by which a non-local product of 
fields is 
expanded into a product of local fields. The utility of the OPE in QCD is 
almost 
immediately evident. The OPE provides us with a basis for dividing processes
with soft 
and hard 
momenta into noncalculable and calculable parts. In QCD it is just 
what we need! The 
running of the strong coupling constant allows us to use perturbation 
theory for 
ultraviolet parts of diagrams, while we hide our ignorance of infrared dynamics
in condensates.

To learn the essentials of the OPE, it is sufficient to consider a simple 
problem in a 
simple theory. Everyone and his brother knows how to calculate the 
Green's function in scalar
$\phi^4$ theory. Below, we will calculate it, and its first correction, with 
normal 
perturbation theory, side by side with the same procedure using  
Wilson's OPE. The 
example is stolen straight out of ref.
\cite{CAN}.
Basically, Wilson's OPE is just the following relation:
\bb
i\int e^{iqx} dx T[j_A(x) j_B(0)] \rightarrow \sum_{n=1}^{\infty} 
D_n^{AB}(q)O_n(0).
\ee
In the example below, we will implement the relation, see how it 
generalizes the normal 
perturbation methods, and directly see its beautiful utility in applications 
to strong 
coupled theories. For these illustrative purposes, we will consider a simple 
example: the 
Higgs model. We can assume the mass squared parameter of the Lagrangian has 
positive sign 
- there are no nonperturbative dynamics - and write the Lagrangian as:
\bb
L= \frac{1}{2}(D_{\mu} \phi)^2 - \frac{1}{2} m_0^2 \phi^2 - 
\frac{\lambda_0}{4!} \phi^4
\ee
where $m_0$  and $\lambda_0$ are the bare mass and coupling constants, 
and the 
field $\phi$ is real. 
Our goal is to calculate the Greens function, $D(q)$, of the field $\phi$ in 
the one-loop 
approximation using traditional methods, $D^{pert}$ and the OPE, $D^{OPE}$.

%\begin{figure}
%\vspace{2.4cm}
%\special{psfile=OPE2.ps hscale=70 vscale=70
%hoffset=-10 voffset=-243 }
%\caption{
%The expansion of the full propagator to $O(\lambda$ in $\phi^4$.}
%\end{figure}

\begin{figure} 
\centerline{\epsfysize=6cm\epsfbox{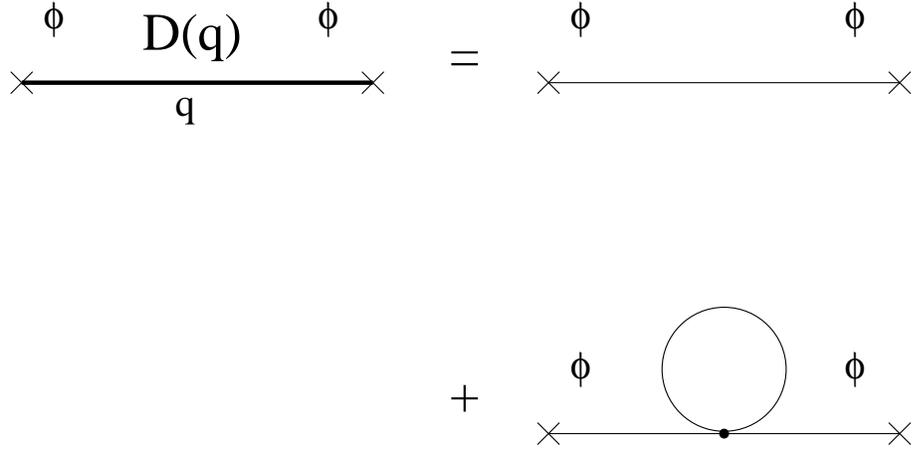}} 
\caption{The propagator for $\phi^4$ theory to $O(\lambda)$.} 
\end{figure}

Consider figure 2.1.
Using the Pauli-Villars regularization method, $D(q)$ can be written as,
\bb
D^{pert}(q) = \frac{1}{q^2-m_0^2} [1 + \frac{\lambda_0}{32 \pi^2}
\frac{M^2 - m_0^2 \; {\rm{ln}} (M^2/m_0^2)}{q^2-m_0^2}].
\label{DPERT}
\ee
Our task is to reproduce this result using the OPE. Essentially, the 
procedure is the 
same as the traditional calculation, except that the perturbative integrals - 
now the coefficients of 
given operators - are 
integrated down to some infrared scale $\mu$, i.e. the perturbative 
graphs containing lines 
with momenta {\it{less}} then $\mu$ flowing through them are `cut'. 
These cut graphs are 
nothing else than the operators. This particular example is an instructive 
one, since we can calculate 
the operators explicitly (unlike in QCD). 

The expression from the OPE we will need for correspondance is:
\bb
D^{OPE}(q) = <0| C_1(\mu) {1} + C_{\phi^2} \phi^2(\mu)|0>
\ee
Here, and in normal applications, the operator $C_1$ stands for all of 
perturbation 
theory - with the caveat that we exclude infrared contributions. The 
excluded infrared 
contribution for $C_1$ rears its head in the subsequent operator terms of 
the OPE expansion, each 
of which has its own perturbative contribution, ad infinitum.
Note another key feature of the OPE - one which we utilize in heavy quark 
applications. The series 
of operators is one of increasing dimension, thus our coefficients must 
come 
along with decreasing dimension. The utility in QCD applications is that the
expansion will be in powers 
$\sim \Lambda_{QCD}/Q$ where $Q$ is some large external momenta. 
To wit, we expand our previous result, eq. \ref{DPERT}, for 
large $q$ and 
thus have a prepared result in normal perturbative methods we can 
compare to our 
upcoming OPE calculation. Expanding for $q^2=-Q^2, Q^2>>m^2$,
we get:
\bb
D^{pert}(q) \rightarrow \frac{1}{Q^2} + \frac{1}{Q^4}[m_0^2 + 
\frac{\lambda}{32\pi^2}
(M^2 - m_0^2 \; {\rm{ln}} \frac{M^2}{m_0^2})].
\label{DPERTEXP}
\ee

Now lets get the OPE result. As stated above, the coefficient $C_1$ is 
nothing but the perturbative 
result with the infrared cutoff $\mu$ - note here that the ease with which 
we 
implement the cutoff $\mu$ in general doesn't exist, for example a 
prescription for 
cutting off infrared dynamics for multi-loop processes would require great 
care (but is in fact possible \cite{TWOLOOP}). In the 
case here, the result is easily achieved - just subtract from the previous 
result, eq. \ref{DPERTEXP} the same 
expression with the ultraviolet regulator M 
replaced by $\mu$. Then, $C_1$ is just
\bb
C_1 = -\frac{1}{Q^2} + \frac{1}{Q^4} [m_0^2 + \frac{\lambda}{32\pi^2}(M^2 - 
\mu^2 - 
m_0^2 \;
{\rm{ln}} \frac{M^2}{\mu^2})]
\label{C1}
\ee
Now we need to 
calculate the remaining piece. The coefficient, $C_{\phi^2}$ is just
figure 2.2a.
\begin{figure} 
\centerline{\epsfysize=6cm\epsfbox{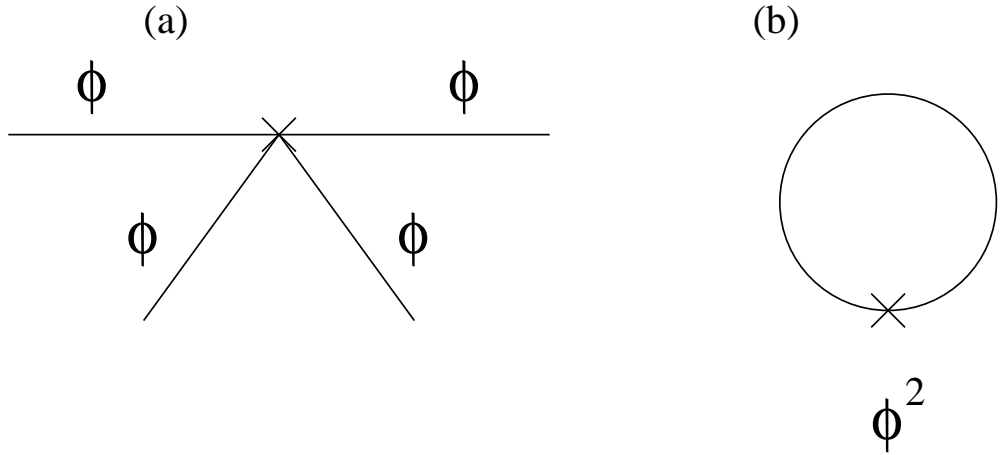}} 
\caption{Figures corresponding to the OPE $\mu$ cutoff prescription for 
the calculation of the propagator in $\phi^4$ theory.} 
\end{figure}
Calculating  the diagram of figure 2.2a yields the result
\bb
C_{\phi} = \frac{\lambda_0}{2 Q^4}.
\ee
where the factor two comes from combinatorics.
Then all we need is the vacuum expectation value $<\phi^2>$. It is just the 
bubble graph of figure 2.2b with the 
entire contribution coming from perturbation theory.
Note, here we remember 
to only integrate in the soft region from 0 to $\mu$, and so,
\bb
<0|\phi^2|0> = \frac{1}{16\pi^2}(\mu^2 - m_0^2 \; {\rm{ln}} 
\frac{\mu^2}{m_0^2}),
\ee
Combining 
these `OPE' equations, we see that they give the same result as the standard 
perturbative methods, i.e.
\bb
D^{OPE} = D^{pert}.
\ee

Turning to QCD, one quickly sees the utility of the OPE. The OPE allows us 
to use 
perturbation theory in the UV - where $\alpha_s$ is small - and get 
some information from perturbative coefficients, and in addition, allows us 
to at least give our ignorance 
of the soft dynamics a name. Here we were able to calculate the 
operators 
directly. In a strongly coupled theory like QCD, we have no idea what the 
Greens 
functions of quarks and gluons are for soft momenta and thus have no 
right to calculate 
them - instead we {\it{extract}} them phenomenologically, as was done 
first in SVZ sum 
rule applications \cite{SVZ}. 
To be a bit more exact about actual practice, 
what is done in QCD is not exactly 
what is done in the example above. Technically, the coefficient functions - 
for infrared safe quantities - are integrated down to $\mu=0$. In the case 
of the D 
meson, it is just because we will borrow the the perturbative results of QED 
(muon decay).  This is 
a blatent double 
count, but as it turns out a justified numerical approximation that works 
well in QCD applications. For instance, in the case of the gluon 
condensate, the true nonperturbative phenomenologically extracted result 
is 
numerically much greater than the perturbatively calculated contribution 
 - a sign of the strongly 
fluctuating vacuum fields. This assumption goes under the name of the 
practical OPE \cite{CAN}. 
The actual, proper, implementation of the OPE would result in the infrared 
contribution of 
coefficient functions being excluded, and thus the introduction of terms 
$(\mu/Q)^n$ (see eq. \ref{C1}). The practical OPE, then, is the 
approximation wherein these 
terms are neglected - an approximation to be checked in each and every
application.

\chapter {The Semileptonic D lifetime - $\Gamma(D)$ - an Introduction}

\begin{center}
``It is so easy to calculate it that it is impossible to make a 
mistake'' - Misha Shifman
\end{center}

If the D meson was just a point particle, all c quark, so to speak, then its 
decay 
wouldn't 
be all that interesting. 
Aside from its CKM matrix element, in fact, it would be the same calculation 
as for the decay of the muon, now a textbook favorite, e.g. \cite{OKUN}. 
Indeed, the semileptonic decay of the muon yields the decay width:
\bb
\Gamma(\mu\rightarrow e \bar{\nu}_e \nu_\mu)= 
\frac{G_F^2 m_{\mu}^5}{192 \pi^3} (1+A^{(1)} \alpha_e)
\ee
where the first order perturbative correction, $A^{(1)}$, was first 
calculated in ref. {\cite{PERT}}.

The calculation for the free quark decay $c \rightarrow s l \nu$ 
goes in the same
way, since the weak currents governing the decay have the same form. Thus
by multiplying 
the above expression by the CKM element squared, $|V_{cs}|^2$, 
we have the D decay width with one loop corrections. 
We can play a quick numerical game and see 
whether the expression agrees with experiment. Setting $\alpha_s = 0$ for now,
\bb
\Gamma(D) = \frac{G_F^2 |V_{cs}|^2 m_c^5}{192 \pi^3}
= 1.1 \times 10 ^{-13} \; \rm{GeV}
\ee
(with\footnote{This value of $m_c$ is taken from 
the early sum rule estimates \cite{SVZ}. I will return later to a full 
discussion of the heavy quark mass.}) $m_c= 1.4\; {\rm{GeV}}$ which is to be 
compared to the experimental 
value: \bb \Gamma_{expt}(D) = 1.06 \times 10^{-13} \; \rm{GeV}.
\ee
It looks like a great agreement! 
We shouldn't be so hasty, however. Corrections to the width will come both from
perturbative and nonperturbative sources.
Unfortunately, as we will see, the result 
gets spoiled by 
these corrections
\footnote{As an aside, note that there is a phase space suppression coming from
the fact that the s quark mass in the final state is massive.
The suppression to the partonic width is given by 
\bb
\phi[z] = 1 - 8 z + 8 z^3 -z^4 + 12 z^2 {\rm{ln}} [z]
\ee
Numerically, taking the s quark mass as around $150 \; {\rm{MeV}}$, 
this effect is
roughly 5 percent, if we are after an accuracy greater than this, then, of
course, these phase space suppressions must be taken into account.}.

To calculate the 
nonperturbative corrections, it will be 
necessary to introduce an additional formalism, first proposed by 
Schwinger in QED \cite{SCHWINGER}, 
and later discovered independantly in QCD in 
\cite{SCHWINGERQCD} with the first applications appearing in 
\cite{ARK} and
\cite{SCHWINGERQCDAPP}. 
For now we focus on the perturbative corrections.  
$A^{(1)}$, the coefficient of the $O(\alpha_s)$ 
corrections can be computed by considering diagrams of the type in fig 
5.1 with gluon brehmsstrahlung occuring off of the heavy quark line or 
the intermediate quark.
The result is 
\bb 
A^{(1)}=  -\frac{2}{3 \pi} (\pi^2 - \frac{25}{4})
\ee
If we wish to borrow this result 
for the QCD decay, we must address two issues. 
First, since the result was first 
calculated in QED, there was no reason to introduce the 
scale $\mu$ separating hard and soft effects, thus we need 
some theoretical justification for using the QED result in QCD where 
there is an infrared problem. 
Second, numerically it is an important issue 
as to what point we normalize $\alpha_s$.

Formally, using the QED result, we imply the seperation 
scale $\mu=0$. Thus using the result in QCD, it means 
literally that {\it{everything}} to $O(\alpha_s)$ 
has been calculated - 
our work is done. Of course, this is nonsense - we have no right 
to believe that the quark/gluon Green's functions at 
low virtuality correspond to the strongly coupled dynamics that 
actually occur for scales less than  
$\Lambda_{QCD}$. 
On the other hand, we don't want to throw away this useful result. 
The 
answer to the puzzle was discussed in 
section 2: we use what is known as the practical OPE {\cite{CAN}}. 
Proper use of the 
OPE requires that we cut our 
integrals off at the infrared scale $\mu$. Doing so would introduce 
corrections of the order $(\mu/Q)^n$ (see eq.(\ref{C1})). 
In the practical OPE,  these terms are discarded. 
The assumption is that the 
perturbative piece that we have no right to include 
(from virtualities 0 to $\mu$) is numerically much smaller than 
the phenomenologically extracted operator. 

As for the second question, what scale do we use for 
$\alpha_s$, there seem to be two options. 
Formally, 
we are allowed to 
take whatever scale of $\alpha_s$ we like since the difference is in the next 
order of $\alpha_s$.  By choosing the appropriate scale, however,
we minimize the  corrections at next order.
The most physically 
reasonable scale is obviously the scale 
$\mu ={\rm{const.}} \times m_c$. 
This is the scale involved in the physical decay - any 
emitted gluon would carry this momenta. 
To see which scale actually comes into play one has to consider next 
order corrections. 
Here, one can use what is called the BLM approach \cite{BLM}. This 
generalized approach is applicable when one deals with a single gluon 
line (dressed by all bubbles). It amounts to inserting the unexpanded 
expression for the running coupling constant 
\bb
\alpha_s(k^2) \approx \frac{4\pi}{b {\rm{ln}} (k^2 \Lambda^2_{QCD})}
\ee
inside the integrand of a one-loop Feynman graph which depends on the 
gluon momentum k, with the subsequent integration over k.
Letting 
$\alpha_s$ run 
we have perturbative integrals like
\bb
\sim \int \frac{\alpha_s(k^2)}{k^2}.
\ee
For the kinematics of the problem at hand,
such integrals will be saturated at scales $m_c$.

Now we would like to 
include nonperturbative effects. 
The most efficient way to calculate the nonperturbative corrections is by using
the background field technique and the Fock-Schwinger gauge. In the next
section, I review both formalisms.

\chapter{Calculations in external fields in QCD}
\begin{center}
{``Getting the damn 2's and $\pi$'s in the right place is the whole point!'' - 
Feynman}
\end{center}
Here, we review a formalism proposed by Schwinger \cite{SCHWINGER}
for the description of 
motion of 
particles in external fields. I will rely heavily on  the formalism 
(developed in \cite{BOOK}) to calculate the 
nonperturbative corrections to the semileptonic D lifetime. 
The Schwinger operator approach is formulated in coordinate space.
Let us introduce a set of states $|x>$ which are the eigenvalues of the 
coordinate 
operator $X_\mu$:
\bb
X_\mu |x> = x_\mu|x>.
\ee
We also introduce the momentum operator $P_\mu$ which satisfies the 
commutation 
relations:
\bb
[P_\mu,X_\nu]=ig_{\mu\nu}
\ee
\bb
[P_\mu,P_\nu] = i g T^a G^a_{\mu\nu}
\ee
where $T^a$ is the color group generator, related to the Gell-Mann matrices 
by $T^a = 
\lambda_a/2$.
In addition to the above relations, we will work in a specific gauge - the 
Fock-Schwinger gauge.
The Fock-Schwinger gauge is the condition 
\bb
x_\mu A_\mu = 0.
\ee
There are two conveniences resulting from this gauge choice that we 
will take advantage of. The first is that the potential is written in terms of 
the field 
strength tensor, $G_{\mu\nu}$, and thus easily yields gauge invariant
expressions. The second is that $A_\mu^a(0) = 0$, as we will now see.

\section{The gauge potential in the Fock-Schwinger gauge}
With our choice of gauge in hand, lets write out an expansion for the 
potential, $A_\mu$. We start with the 
identity:
\bb
A_\mu^a(x) = \int_0^1 \alpha d\alpha G^a_{\rho\mu}(\alpha x) x_{\rho}.
\ee
We can prove it, since, from our gauge condition,
\bb
A_{\mu}(y) = \frac{d}{dy_{\mu}} (A_{\rho}(y) y_{\rho}) - y_{\rho} 
\frac{dA_{\rho}(y)}{dy_{\mu}}= - y_{\rho} \frac{dA_{\rho}(y)}{dy_{\mu}},
\ee
then inserting the definition of the field strength tensor,
\bb
y_{\rho} \frac{dA_{\rho}(y)}{dy_{\mu}}=y_{\rho}G_{\mu\rho} + y_{\rho} 
d_{\rho} A_{\mu} 
+ y_{\rho} [A_{\mu},A_{\rho}]=y_{\rho}G_{\mu\rho} + y_{\rho} d_{\rho} 
A_{\mu}, 
\ee
so then 
\bb
A_{\mu}(y) + y_{\rho} \frac{dA_{\mu}(y)}{dy_{\rho}} = y_{\rho} 
G_{\rho\mu}(y).
\ee
Now, reparametrize $y=\alpha x$,
then
\bb 
\frac{d}{d\alpha}(\alpha A_{\mu}(\alpha x) = A_{\mu}(\alpha x) + \alpha 
\frac{d}{d \alpha} 
A_{\mu}(\alpha x),
\ee
and so,
\bb
\int^1_0 \alpha x_{\rho} G_{\rho\mu} (\alpha x) d\alpha = 
\int^1_0 \frac{d}{d\alpha}(\alpha A_\mu (\alpha x) d\alpha
= x_\mu A_\mu .
\ee

Now, to get the expansion for the gauge potential,
write the field strength tensor as,
\bb
G_{\rho\mu}^a (\alpha x) = G_{\rho \mu}(0) + \alpha x \frac{d}{d(\alpha 
x)} 
G_{\rho\mu}(0)+...
\ee
insert this into the integral representation and note that since in this 
gauge,
$d_{\alpha} = D_{\alpha}$
\bb
A_\mu^a(x) = \frac{1}{2\cdot 0!} x_\rho G_{\rho\mu}(0) + \frac{1}{3\cdot 1!}
x_\alpha x_\rho (D_\alpha G_{\rho \mu}(0)) + ....
\label{GE}
\ee

\section{Calculating propagators in the Fock-Schwinger gauge}
As an 
exercise in this gauge, we can try 
and calculate propagators using the external field method \cite{ARK}. 
We will need such propagators later. 
The scalar propagator 
for a 
massless particle in the Fock Schwinger gauge is written as,
\bb
S(q) =
\int dx e^{iqx} <x|\frac{1}{P^2}|0>=\int dx <x|\frac{1}{(P+q)^2}|0>,
\label{scal}
\ee
since
\bb
e^{iqX} P_{\mu} = (P_\mu + q_\mu) e^{iqx}.
\ee
Now expand this propagator in 
powers of $P/q$. Here we use a simplification 
resulting in our choice of the Fock-Schwinger gauge referred to above:
\bb
A_\mu|0> = 0.
\ee
Thus, our main strategy is to expand in $P/q$ and pull all gauge potentials 
to the right. Also, since
\bb
\int dx <x|p_\mu \cdot \cdot \cdot = \int dx dy i d_\mu \delta(x-y)<y|\cdot
\cdot \cdot |0> =0
\ee
we pull all p's to the left.
The only other ingredient is just commutators between p's and A's, which are
trivial to evaluate in the Fock-Schwinger gauge.
Expanding equation (\ref{scal}) in $P/q$,
\bb
\frac{1}{(P+q)^2} = \frac{1}{q^2} - \frac{2Pq}{q^4} + \frac{4(Pq)^2}{q^6} - 
\frac{P^2}{q^4}
+ \frac{P^2(2Pq) + (2Pq)P^2}{q^6} - \frac{8 (Pq)^3}{q^8} + O(P^4)
\ee
and performing the procedure described above, we get
\begin{eqnarray}
D(q) & = & \frac{1}{q^2} - \frac{g}{3 q^6} D_{\alpha} G_{\alpha \rho} q_{\rho}
\nonumber \\
&\;& - \frac{g^2}{2 q^8}(q_\alpha G_{\alpha \rho} G_{\rho \gamma} q_{\gamma} +
\frac{1}{4} q^2 G_{\alpha \rho} G_{\alpha \rho}) - \frac{ig}{q^8}q_\gamma
D_\gamma D_\alpha G_{\alpha \rho} q_\rho + ...
\end{eqnarray}
The case of the scalar propagator is 
of course easiest since we do not have to worry 
about the 
anticommuting $\gamma$ matrices. We will, however, need the result for spinor
particles. The result\footnote{Note, here I use $\hat{p}$ instead of the 
Feynman
convention, $p \! \! /$, to denote
the Dirac matrices - Misha Voloshin taught me this, explaining that to him,
a slash meant that the given variable was crossed out!}
 (in the massless case) of this calculation is, to 
$O(DG):$ 
\begin{eqnarray} 
S(q)&=&\frac{1}{\hat{q}} - \frac{g}{2q^4} q_{\alpha} 
\tilde{G}_{\alpha\rho} \gamma_\rho \gamma_5 
\nonumber \\ & \; & + \frac{g}{3 q^6} 
[q^2 D_\alpha G_{\alpha\rho} \gamma_{\rho} - \hat{q} 
D_\alpha 
G_{\alpha\rho} q_\rho - q_\gamma D_\gamma q_\alpha G_{\alpha \rho} 
\gamma_\rho \nonumber \\ & \; & - 
3 i q_\gamma D_\gamma q_\alpha \tilde{G}_{\alpha\rho} \gamma_\rho 
\gamma_5].
\end{eqnarray}

\chapter{Nonperturbative corrections to the lifetime of the D}
\begin{center}
``The heavy quark is heavy - its like a locomotive flying through mosquitoes'' 
- Misha Voloshin
\end{center}
In section 3, we used the textbook calculation for muon decay to get our hands 
dirty for the case of the 
D meson. 
Let me derive this result in a fairly non-standard but equivalent 
way by using the 
optical theorem. The technique will be a bit long-winded for the standard 
partonic calculation, but 
will simplify life greatly when we are after nonperturbative corrections.

\section{Derivation of parton result using the optical theorem}
The optical theorem relates the imaginary part of a forward scattering 
amplitude to its observable cut process. The optical relation
valid for the total widths is,
\bb
\Gamma = \frac{1}{M_D} {\rm{Im}} <D|\hat{T}|D>,
\ee
(the relation above implies the use of relativistic normalization of states)
which is our starting point.
Here $\hat{T}$ is the time ordered product of, in this case, the two weak 
currents 
governing the semileptonic decay. From figure 5.1,
\begin{figure} 
\centerline{\epsfysize=6cm\epsfbox{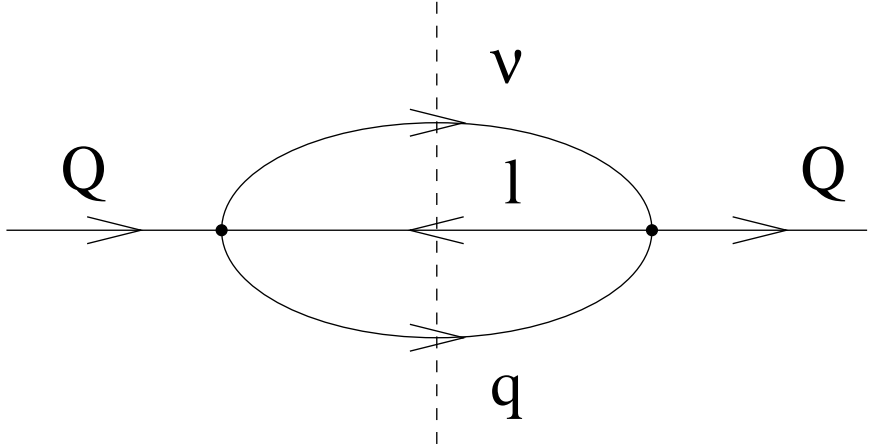}} 
\caption{The forward scattering diagram for semileptonic decay. THe 
dashed line represents the `cut', i.e. the imaginary part of the diagram} 
\end{figure} 
we get the transition operator in configuration space as,
\bb
\hat{T} = -\frac{G_F^2}{2}\int d^4x \bar{Q}(x) \Gamma_\mu S_q (x,0) 
l_{\mu\nu}(x)
\Gamma_\nu Q(0),
\label{THAT}
\ee
where Q represents the heavy quark external line, $S_q$ is the light quark 
propagator, 
the $\Gamma$'s are the usual $V\times A$ Dirac matrices ($\Gamma_\mu = 
\gamma_\mu (1-\gamma_5))$, and the lepton tensor is,
\bb
l_{\mu\nu}(x) = {\rm{Tr}} [\Gamma_\mu \frac{-1}{2 \pi^2} 
\frac{\hat{x}}{x^4}\Gamma_\nu 	
\frac{1}{2\pi^2}\frac{\hat{x}}{x^4}]
= - \frac{2}{\pi^4}\frac{1}{x^8} (2 x^\mu x^\nu - x^2 
g^{\mu\nu}).
\ee
To get the parton result, take the light quark propagator as free:
\bb
S_q(x,0) = \frac{\hat{x}}{2 \pi^2 x^4}.
\ee
Upon convolution of Dirac matrices, $\hat{T}$ becomes:
\bb
\hat{T} = \int d^4x \bar{Q}(x) f(x) (1+\gamma_5) Q(0),
\label{T}
\ee
where $f(x)$ is a function of $x$.
The interaction of the heavy quark with the light degrees of freedom enters
through the Dirac operator, $D_\mu = d_\mu - igA_\mu^a t^a$. The background
gluon field $A_\mu$ is small compared to $m_Q$, and thus there is a large
`mechanical' part in the x-dependance of $Q(x)$
\cite{MECH}:
\bb
Q(x) = e^{im_Qt} \tilde{Q}(x).
\ee
Our transition operator is still a nonlocal object.
To write 
it as a local object we should Taylor expand the 
heavy quark field in the Fock-Schwinger gauge:
\begin{eqnarray}
\partial
\tilde{\bar{Q}}(x)&=& (1+ x_{\alpha 1} \partial_{\alpha 1} + \frac{1}{2} 
x_{\alpha_1}
x_{\alpha_2} \partial_{\alpha 1} \partial_{\alpha 2}+...) \tilde{Q}(0) 
\nonumber \\ &=&(1+x_{\alpha 1} 
D_{\alpha_1} + \frac{1}{2}
x_{\alpha_1} x_{\alpha 2} D_{\alpha 1} D_{\alpha 2}+...) \tilde{Q}(0).
\end{eqnarray}
The above expansion is legal since in the Fock-Schwinger gauge ordinary 
derivatives may be substituted by full derivatives at the origin.
We can get rid of the $\gamma_5$ term in equation (\ref{T}) since,
through the equations of motion it appears as a full derivative
\bb
<D|\frac{\partial}{\partial x_\mu} (\bar{Q} \gamma_\mu \gamma_5 Q)|D> = 2 i 
m_Q <D|\bar{Q}\gamma_5 Q|D>,
\ee
and the
derivative of the forward scattering matrix element gives zero (this matrix
element and others like it containing $\gamma_5$ are zero from parity arguments
anyway).
Upon convolution of the lepton tensor with $\Gamma_\mu$'s, etc., we are left 
with 
\bb
\hat{T}={\rm{const.}}\int d^4x e^{ipx} \tilde{\bar{Q}}(0) \frac{\hat{x}}
{x^{10}}Q(0).
\ee
To do the integral (see the Appendix in \cite{BOOK}), we use the result
\bb
I= \int d^4 x \frac{e^{ipx}}{(x^2)^n} = \frac{i(-1)^n2^{(4-2n)}\pi^2}{\Gamma(n-
1)\Gamma(n)}
(p^2)^{n-2} {\rm{ln}}[-p^2],
\ee
which gives, in our case,
\bb
\hat{T}= -i\gamma_\mu \frac{d}{dp_\mu} I 
= -i\gamma_\mu \frac{d}{dp_\mu}\frac{i(-1)^5 2^{-6} \pi^2 p^6 ({\rm{ln}}
|p^2| + i\pi)}{4! \; 3!}
\ee
and so finally\footnote{Note, this, the near end result for the parton 
decay width, is the
beginning of where we
look for the nonperturbative $1/m_c^2$ pieces.} we get \bb
\hat{T} = i\frac{G_F^2 V_{cs}^2 \bar{Q}(0) p^4 \hat{p} Q(0)}{384 \pi^3}.
\label{START}
\ee
To leading order, $p^4 = \hat{P}^4$ (we will see it later). 
Substituting this result, and using 
the Dirac equation, we are left with
\bb
\Gamma_0(D) = \frac{1}{M_D} {\rm{Im}} (\frac{i G_F^2 V_{cs}^2 m_c^5}{384 \pi^3})
<D|\bar{Q}(0) Q(0)|D>
\ee
Also to leading order (we will see it explicitly later (equation (\ref{QQ})), 
\bb
<D|\bar{Q}(0)Q(0)|D> \sim <D|\bar{Q}(0) \gamma_0 Q(0)|D> = 2 M_D,
\ee
i.e. this operator merely counts the number of c quarks in the D meson - it 
is basically the number 
density.  Here we get 
into points 
concerning $\mu$ dependence of operators involved in our expression. 
Technically, 
$\bar{Q} Q$ carries $\mu$ dependence. Moreover, if $\mu$ is 
large, say a 
few GeV, then we are not entitled to say that there is only one c quark in 
the D - we 
must evolve the operator down to a few hundred MeV - a typical hadronic 
scale. Later we will see that the heavy quark mass should be at a 
normalization point $\mu=m_c$. If so, then the operator $\bar{Q}Q$ is 
also at $\mu = m_c$. We must evolve this operator down to a 
lower normalization point where we can say that all fluctuations are soft, 
and thus do not involve the creation of heavy quark anti-heavy quark pairs. 
Evolving this operator down is equivalent to pumping the fluctuation 
contained in the operator into the coefficient function. These 
fluctuations are actually already contained in the perturbative 
correction, $A^{(1)}$ anyway, so we can forget about them. Now, upon
combining 
the relations above,
we finally get the expected result:
\bb
\Gamma_0(D) = \frac{G_F^2 V_{cs}^2 m_c^5}{192 \pi^3}.
\ee

\section{$O(1/m_Q^2)$ corrections}
To get the $O(1/m_Q^2)$ corrections, 
we will need one additional piece of machinery, the expansion
\bb
\bar{Q}Q = \bar{Q}\gamma_0Q - \frac{\bar{Q}[(i\vec{D})^2 - (i/2)\sigma 
G]Q}{2m_Q^2}
+O(1/m_Q^4)
+{\rm {total \; derivatives}}.
\label{QQ}
\ee
The $O(1/m_Q^2)$ result in the above equation was first derived in
\cite{BUV}.
The relation itself  
can be easily established
from the equations of motion. 
Since,                                               
\bb
\frac{1-\gamma_0}{2} c = \frac{1}{2 m_c} \hat{\pi} c
,
\ee
then, 
\bb 
\bar{c} \frac{1-\gamma_0}{2} c = \bar{c}
\frac{1-\gamma_0}{2}\frac{1-\gamma_0}{2} c = \frac{1}{4 m_c^2} \bar{c} \hat{p}
\hat{p} c,
\ee
which implies that
\bb 
\bar{c}(1-\gamma_o)c = \frac{1}{2 m_c^2} \bar{c} (\pi^2 + \frac{i}{2}
\sigma G)c.
\ee
Then, using the equations of motion again, we see that $\bar{c} \pi_0^2 c$
actually reduces to an operator of dimension 7, since
\bb
\pi_0 Q = - \frac{\pi^2 + (i/2) \sigma G}{2 m_Q} Q
\ee
and so we arrive at the 
result above.

There are various possible sources for $1/m_Q^2$ corrections to the parton
result. First, recall that the first term of the gauge potential in the
Fock-Schwinger gauge is $\sim G$.
This term actually gives a contribution of $O(1/m_Q^3)$
(we will see these terms later). 
There are however two other sources of $1/m_c^2$ terms. 
The first is from equation (\ref{QQ}) 
above in the expansion of $\bar{Q}Q$, 
and the 
second is from the identity
\bb
P^2 = \hat{P}^2 - \frac{i}{2} \sigma G
\ee
which we plug into eq. \ref{START}
(the identity above is obvious since $\gamma_\mu \gamma_\nu = g_{\mu \nu} + 
\sigma_{\mu \nu}$). 
Using these two results, we then plug $\hat{T}$ into the optical relation. 
The result is
\bb
\Gamma(D) = \frac{G_F^2 V_{cs}^2}{192 \pi^3} m_c^5 \times [1 +A^{(1)} \alpha_s 
- \frac{3}{2} \frac{\mu_G^2}{m_c^2} -\frac{1}{2} \frac{\mu_\pi^2}{m_c^2}].
\ee

Two points are in order here. First, notice that there is no contribution
coming from operators $O(1/m_c)$, i.e. there is no operator of dimension 
4 \cite{CGG}. 
Second, since $\Gamma$ is a Lorentz scalar, we should not be surprised to
see that the only operators appearing in $\Gamma$ are Lorentz scalars
($\mu_\pi^2$ enters only in combination with $Q\gamma_0 Q$ - from the original
Lorentz invariant operator $\bar{Q}Q$).

Our task is now somewhat complete - we have the partonic width, and the leading
perturbative and nonperturbative 
corrections. 
Now we would like to do some numerics and see whether we match experiment. 
This is an area of hot debate for 
sure, but as we will see, the failure of the perturbative and 
nonperturbative corrections to give the experimental result is fairly 
insensitive to most of the numerical debates.
Clearly the most important numerical issue will be the mass of the c 
quark, since it 
appears as the fifth power in the parton result. 
We now turn to a complete discussion of numerics.

\chapter{Theoretical meaning and numerics of nonperturbative 
parameters}
\begin{center}
``...the set of values to be passed from the elders to the young 
generations included the idea that high energy physics is an experimental 
science that {\it{must}} be very closely related to phenomena taking 
place in nature....'' - Misha Shifman
\end{center}

\begin{center}
``So Dave, you would like a problem to work on? Well, I must warn you 
that I will not give you something very mathematical, but instead more 
phenomenological - any results you will obtain will probably have direct 
relevance to current experiments..." - Misha Shifman, much to my excitement!
\end{center}

In this section, I discuss the meaning and numerical values of the heavy 
quark mass $m_c$, $\mu_\pi^2$, and $\mu_G^2$. 

\section{The meaning of the heavy quark mass}
The issue of the heavy quark mass can easily be anticipated to be a sticky 
one. As 
everyone knows, quarks are not color singlets and thus do not appear 
isolated in nature (quarks are not asymptotic states, if you like). 
And yet, the theory of QCD is built upon quarks and gluons. Clearly, a key 
parameter in 
any QCD calculation will then be the quark mass - it enters in the QCD
Lagrangian. 
In the calculation of the D 
meson lifetime, we will be borrowing perturbative results from QED: the 
one loop 
correction to muon decay. In the QED calculation, the muon mass used is 
just the pole 
mass. In QCD, this becomes a problem since by using the pole mass of the 
heavy 
quark, we are unjustifiably including long distance dynamics. Using the 
pole mass, we 
are essentially assuming that we have knowledge of gluonic Greens 
functions at all 
scales - a clearly unjustified assumption. In fact, it can be shown that the 
pole mass 
of the heavy quark is defined through an asymptotic series in $\alpha_s$, 
leading to an 
ambiguity of order $\Lambda_{QCD}$. 
This is a strong statement, and with the 
propagation of the 
words `pole mass' in almost every paper on QCD, one gets the feeling 
that this fact is somehow avoided, or misunderstood. The 
misunderstanding 
comes from whether we are talking about the `pole mass' or the `one loop 
pole mass'. Although, as we will see below, the pole mass is ill-defined, 
one can rightfully use the one loop 
pole mass, as I will below, in any typical problem. In the problem of the D 
meson decay, 
use of the one-loop pole mass really helps simplify life - we can just use 
the well-known 
QED result for the muon decay, without including in it some scale 
dependance (i.e. cut 
off loop integrations at some scale). Then, use of the pole mass is proper, 
since we 
have, in effect, set the scale - at zero virtuality - by using the QED result. 
Here, we see 
the conceptual difference between the $\phi^4$ exercise offered earlier, 
and what is 
called the practical OPE by QCD practitioners. Let me then, in turn, review 
two main points concerning the heavy quark mass:

\subsection{Illegetimacy of the pole mass}
Although commonly used in practical applications, 
it should be understood that the pole mass of the quark is an ill-defined 
object. 
The subject is thoroughly discussed in \cite{POLEMASS}, and I only briefly
touch upon some essentials here.
To see that the pole-mass is ill-defined, 
consider the relationship between the pole mass and the 
running mass,
\bb
m_{pole}^{(k)}=
m_Q(\mu)\sum_{n=0}^k C_n(\frac{\mu}{m})(\frac{\alpha_s(\mu)}{\pi})^n
\ee
The above series is not well behaved, 
diverging factorially. 
Such a series is an asymptotic series, and must be truncated at a 
critical order in the expansion. After truncation, the remainder of the 
series shows up as an uncertainty (knowing the exact remainder would constitute
having knowledge of the vacuum fluctuations). 

\begin{figure} 
\centerline{\epsfysize=3cm\epsfbox{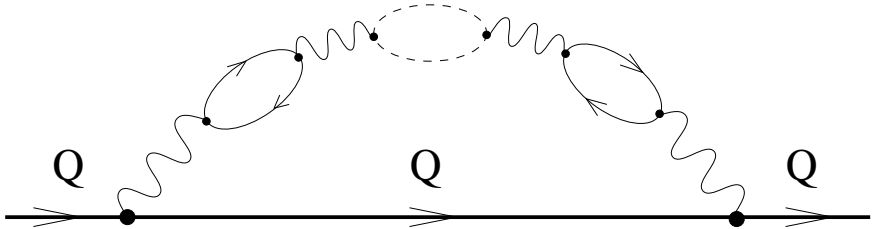}} 
\caption{The diagram corresponding to the mass renormalization due to the 
renormalon gluon bubble chain.} 
\end{figure}

%\begin{figure}
%\vspace{2.4cm}
%\special{psfile=revfig1.ps hscale=70 vscale=70
%hoffset=-10 voffset=-243 }
%\caption{
%Perturbative diagrams leading to the IR renormalon
%uncertainty in $m_Q^{pole}$ of order $\Lambda$.
%The number of bubble insertions in the gluon propagator
%can be arbitrary}
%\end{figure}

One particular source of the factorial divergence is the infrared renormalon.
The infrared renormalon is the effect of the insertion of the bubble 
chain in 
gluon lines, i.e. allowing the running of the strong coupling constant, 
$\alpha_s \rightarrow \alpha_s(k^2)$, see figure 6.1. 
For nonrelativistic momenta, $k^2 << \mu^2$, 
the expression for the mass correction is
\bb
\delta m_Q \sim \int\frac{d^4k}{(2\pi)^4ik_o}\frac{4\pi\alpha_s(-k^2)}{k^2} 
\sim \int \frac{d^3\vec{k}}{4\pi^2}\frac{\alpha_s(\vec{k}^2)}{\vec{k}^2}.
\ee
Expressing the 
running $\alpha_s(k^2)$ in terms of $\alpha_s(\mu^2)$, $k^2<\mu^2$,
\bb
\alpha_s(k^2) = \alpha_s(\mu^2)(1+\frac{\alpha_s(\mu^2)}{4 \pi} b \; {\rm{ln}}
\frac{k^2}{\mu^2})^{-1} \; \; , \; b=\frac{11}{3} N_c - \frac{2}{3}N_f
\ee
one can expand $\alpha_s(k^2)$ 
in a power series of $\alpha_s(\mu^2)$. One then finds for the (n+1)-th order 
contribution,
\bb
\frac{\delta m_Q^{(n+1)}}{m_Q} 
\sim {\alpha_s(\mu^2)}{\pi} n! (\frac{b \; \alpha_s(\mu^2)}{2\pi})^n.
\ee
Observe that the coefficients grow factorially and contribute 
with the same sign\footnote{Same 
sign series can not be summed since the pole of such 
a series will lie on the real axis, and the contour around 
such a pole when Borel summing cannot be unambiguously defined.}.
Therefore one cannot define the sum unambiguously. An optimal truncation 
leaves one with an irreducible error of $O(\Lambda_{QCD})$. 
To see this, note that truncating the factorial 
growing series at optimal order (see the section on duality 
violations in chapter 10) leaves an exponential uncertainty,
\bb
\Delta(m_Q^{pole} - m_Q(\mu_0)) \sim \mu_0 \; 
{\rm{exp}}(-\frac{2\pi}{b \alpha_s(\mu_0)}) \sim \Lambda_{QCD}.
\ee

\subsection{Irrelevance of the pole mass}
Anyway, in applications, 
it is the running mass 
and not the pole mass which enters into our theoretical expressions. 
As explained above, the perturbative series for $m_Q^{pole}$ diverges 
factorially. In our expression for the width, a similar divergence occurs in
the $\alpha_s$ expansion due to renormalons. If the running mass is 
used, however, both effects
combine to cancel each other. This point is investigated thoroughly in
\cite{POLEMASS}. The result is really neat. It tells us, basically, to perform
the following procedure when extracting quantities like the mass from total
widths \cite{KOLYA}. First, write the total width as an expansion in $\alpha_s$
and $1/m_c$. The $\alpha_s$ series is, as in this case, to be taken at 
$\mu=0$,
i.e. the QED results can be borrowed. Next, one can numerically investigate the
coefficients of the $\alpha_s$ series. In the BLM procedure, the corrections
at two-loops are easily calculated, and give large coefficients. Ref.
\cite{POLEMASS} tells us that these coefficients are nothing more than the
signal of the factorially diverging perturbation series. Not to worry, since,
by inserting the running mass, $m_Q(m_Q)$ we can `cancel' the infrared 
renormalon
effect responsible for these large coefficients. After this procedure, 
aside from issues of
duality violation, it is perfectly alright to extract the running mass from the 
theoretical expression for the width. The running mass itself is easily
calculated (see figure 6.2):
\begin{figure} 
\centerline{\epsfysize=3cm\epsfbox{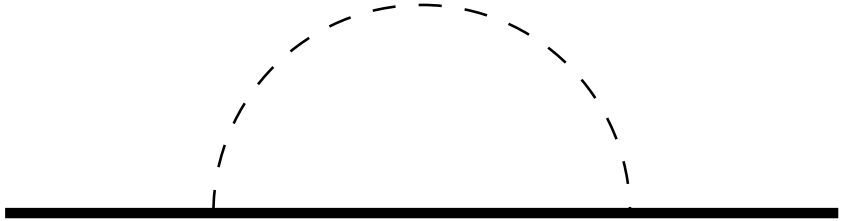}} 
\caption{The diagram corresponding to the running mass of the heavy quark.} 
\end{figure}
to one loop the result is
\bb
m_Q(\mu) = m_Q(\mu_0) + \frac{2 \alpha_s}{3 \pi} (\mu_0 - \mu).
\ee

\section{The value of $m_c$}
In this subsection we address the question `what is the value of $m_c$'.
$m_c$ was first extracted from 
the QCD sum rule application to charmonium, 
and so we review the sum rules here.
As a cross check, I also present an estimate of $m_c$ from $m_b$.
\subsection{$m_c$ from sum rules}
Here, I derive one of the first OPE type applications in QCD - charmonium 
sum rules, \cite{SVZ}.
\begin{figure}
\centerline{\epsfysize=3cm\epsfbox{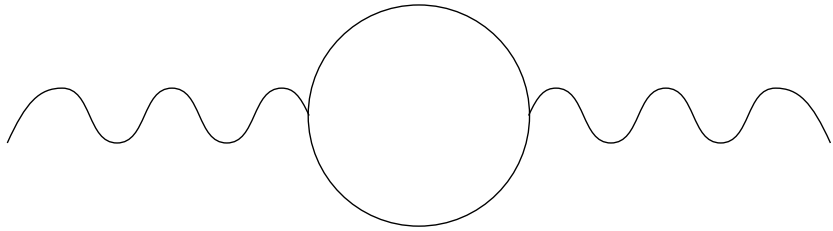}}
\caption{The diagram corresponding to $e^+ e^-$ annhilation.}
\end{figure}
The derivation relies on the standard OPE treatment found in sum rules, or 
heavy quark applications.  
The starting point is quark-hadron duality: on the left-hand side of the 
equation, there is 
a theoretical construction in terms of quarks and gluons, on the right hand 
side is a 
quantity extracted from some hadronic experiment. The assumption of 
quark-hadron 
duality is built in from the first step. As we will see later, this is a well 
justified 
assumption in applications where we rely on dispersion relations  like 
the sum rules\footnote{It is another reason I review charmonium sum
rules.}, 
but we must take care in making the assumption for the D meson.
The reason for this is thoroughly discussed in chapter 10, and resides in 
the fact that sum rules rely on Euclidean kinematics, while the lifetime 
of the D meson is of Minkowskian kinematics.

Our starting point in the investigation is the T product of two 
electromagnetic currents, (see figure 6.3):
\bb
i<0|\int dx e^{iqx}T[j_{\mu}(x)j_{\nu}(0)]|0> =
(q_{\mu}q_{\nu} - q^2 g_{\mu\nu}) \Pi(Q^2),
\ee
where  $j_{\mu} = \bar{c}\gamma_\mu c$, and $Q^2=-q^2$.
The polarization operator, $\Pi$ satisfies dispersion relations:
\bb
-\frac{d}{dQ^2} \Pi = \frac{1}{12\pi^2 Q^2} \int\frac{R_c(s) ds}{(s+Q^2)^2},   
\; \; R_c = \frac{3 s \sigma_c}{4 \pi \alpha^2}.
\ee
We will especially be interested in the moments of $\Pi$,
\bb
M_n = \frac{1}{12 \pi^2 Q^2} \int \frac{R_c(s) ds}{s^{(n+1)}} = \frac{1}{n!}
(-\frac{d}{dQ^2})^n \Pi |_{Q^2\rightarrow 0}.
\ee
The simplest quark loop, is easily shown to be:
\bb
M_n^{(0)} = \frac{3}{4\pi^2} \frac{2^n (n+1)(n-1)!}{(2n+3)!!}\frac{1}{(4 
m_c^2)^n}.
\ee
Of course, the point of the sum rules was to go beyond such a simple 
expression, and include not only $\alpha_s$ corrections, but also new 
power corrections 
like those we have been discussing above. In the case of charmonium, the 
first power 
correction is the gluon condensate. As it turns out, however, the 
contribution of the 
gluon condensate is highly suppressed for low moments, as are 
perturbative corrections - and so the 
key parameter for low moments is just $m_c$.  Extracting it from the data, 
and ascribing a rather large uncertainty \cite{SVZ}, 
\bb
m_c(m_c)= 1.26 \; \pm 0.10 \; {\rm{GeV}}
\ee

\subsection{$m_c$ from $m_b$}
As a check of this result, we can examine another source where $m_c$ can be
extracted. In \cite{FANE} the following result was first derived,
\bb
M_B-M_D=m_b-m_c+\frac{\mu_\pi^2-\mu_G^2}{2 m_b} 
-\frac{\mu_\pi^2-\mu_G^2}{2 m_c}+O(m_c^2,m_b^2).
\ee
The meson masses are measured accurately, ($M_B=5.287 \; {\rm{GeV}}$ and
$M_D=1.864 \; {\rm{GeV}}$), and the value of $m_b$ is known with
high accuracy from upsilon sum rules \cite{VOLMASS}
$m_b(1 {\rm{GeV}}) = 4.64 \pm 0.05 \;{\rm{GeV}}$.
Plugging in these values, we get
$m_c = 1.2 \pm 0.05 \; {\rm{GeV}}$
where the uncertainty comes mainly from the uncertainty in $\mu_\pi^2$ (see 
below).

\subsection{Comments on the literature}
Despite the above arguments, there are still claims in the literature
that one can take the pole mass of the charmed quark to be very
high, $m_c =1.65 \; \rm{GeV}$. 
A typical argument can go as follows.
First, expressions for $\Gamma(D)$, and $m_c^{pole}$ are given, both with 
full one-loop and BLM type two-loop corrections. The explicit expression for
the mass is
\bb
m_c^{pole} = \bar{m_c}[1+\frac{4}{3}\frac{\alpha_s(m_c)}{\pi}(1+1.04)]
\ee
where `1.04' is due to the BLM type two-loop correction. 
The high value $m_c =1.65 \; \rm{GeV}$ 
is obtained by directly accomodating the experimental measurement of
$\Gamma(D)$ with the full theoretical expression, including 
nonperturbative effects. Then one uses the
above equation to run $m_c^{pole}$ to $\bar{m_c}$, and obtain $\bar{m_c} \sim
1.34 \; \rm{GeV}$ which is claimed to be close to the sum rule prediction.

A clear signal that this analysis is in trouble is that the value of the 
pole mass
of the b quark extracted is $m_b^{pole} \sim 5.0 \; {\rm{GeV}}$ in disagreement
with Voloshin's extraction, $m_b^{pole} \sim 4.8 \; {\rm{GeV}}$.
What is wrong with the above analysis? The error is nothing more than a
manifestation of the problems occuring when using the pole mass. Clearly, the
above expansion for $m_c$ is not converging well. This fact is actually to the
benefit of those who would like to fit $m_c^{pole}$  
from $\Gamma(D)_{expt}$ and then
run down to $\bar{m_c}$ - the factor `1.04' allows such a large running.
The problem is that it is not the pole-mass which really enters theoretical
expressions, but the running mass. 

Above, it was discussed that the proper mass
to use in expressions for the full width is the running mass. In ref.
\cite{KOLYA} it was explicitly shown that using the proper running mass, we
reduce large second order corrections to the width in heavy quark
decay. Explicitly, 
ref.
\cite{KOLYA} showed that for the semileptonic width
we have large two-loop corrections,
\bb 
\frac{\Gamma(B)}{\Gamma_0(B)} = 1 + a_1(\frac{\alpha_s}{\pi}) + a_2
(\frac{\alpha_s}{\pi})^2
\ee
where 
$a_2$ is the BLM type correction, and $a_1 = -2.41$, and $a_2 = -19.7$
\footnote{the result for $a_2$ 
is presented in the V scheme, a value around $a_2 \sim -30$
would be obtained for the $\bar{\rm{MS}}$ scheme used in some papers,
allowing an even GREATER value of $m_c$.},
but upon insertion of the proper OPE scale, say, $\mu \sim 400 \; 
\rm{MeV}$ , $a_1 \rightarrow -0.5$ and $a_2
\rightarrow - 0.4$, and thus upon insertion of the running mass, the second
order corrections become small. Upon insertion of these corrections to the
total with, we obtain
\bb
\Gamma(D) = 
\Gamma_0 (1-0.07-0.008-0.27-0.09) = 
\Gamma_0 (0.55),
\ee
where $-0.07,-0.008,-0.27, {\rm{and}} -0.09$ are from $\alpha_s$, 
$\alpha_s^2$, $\mu_G^2$, and $\mu_\pi^2$ corrections respectively (I will 
discuss these numerics in full in the next section).

Now, extracting the mass from this result, the running mass mind you, we get 
\bb
m_c(400 \rm{MeV}) = 1.57 \rm{GeV}
\ee
which is in disagreement with the sum rules estimate, a result which I 
discuss again in section 7. When using the proper OPE procedure, outlined 
above, the extraction of $m_c$ from semileptonic D decay and charmonium 
sum rules disagree.

\section{About $\mu_G^2$}
The mesonic matrix elements of the chromomagnetic operator, $\sigma G$ 
can be extracted from 
the hyperfine splitting.
First, note that 
\bb
\vec{\sigma}\cdot \vec{G} = \vec{\sigma}\cdot \vec{B} + O(1/m_c).
\ee
Now, $\vec{B}$ is an axial vector and is determined by the dynamics of the 
light spectator cloud, thus we have only one choice, $\vec{B} 
= {\rm{const}} \times \vec{S_q}$
and so the matrix element must be:
\bb
<Q\bar{q}| \frac{g \vec{\sigma}\cdot \vec{B}}
{2m_Q}|Q \bar{q}> = C <Q\bar{q}|\vec{S}_Q \cdot
\vec{S}_{\bar{q}}|Q \bar{q}>.
\ee
Now, 
\bb
\vec{S_Q}\cdot \vec{S}_{\bar{q}} = \frac{1}{2} (\vec{S}_{\bar{Q}} 
+ \vec{S}_{q})^2 - \frac{1}{2} (\vec{S}_{Q}^2 + \vec{S}_{\vec{q}}^2)
= \frac{1}{2} S_{TOT} (S_{TOT} + 1) - \frac{3}{4},
\ee
and,
\bb
<0^-| \vec{\sigma}\cdot \vec{B}|0^-> = -3<1^-|\vec{\sigma}\cdot \vec{B}|1^->,
\ee
since for an uncorrellated spin state, the matrix element is zero.
Then, we find simply that,
\bb
M_{(0^-)}- M_{(1^-)} = <0^-|\vec{\sigma}\cdot \vec{B}|0^-> - <1^-|\vec{\sigma} 
\cdot \vec{B}|1^-> = \frac{4}{3}<0^-| \vec{\sigma}\cdot\vec{B}|0^->
\ee
and so, 
\bb
<\mu_G^2>_{D} = <D|( \bar{Q} \frac{i}{2} \vec{\sigma} \cdot \vec{G}) Q|D> 
= \frac{3}{2}m_c 
(M_{D^*} - 
M_D) = \frac{3}{4} (M_{D^*}^2 - M_{D}^2).
\ee
Numerically, taking this mass splitting from the D and B system 
respectively,we get:
\bb
<\mu_G^2>_D = 0.41 \; 
{\rm{GeV^2}}, <\mu_G^2>_B = 0.37 \; 
{\rm{GeV^2}}.
\ee
A measure of the reliability of the nonperturbative expansion is then 
provided by 
$\sqrt{<\mu_G^2>_D/m_c^2} = 0.46$. And thus, right off the bat there is good 
reason to pursue 
higher order nonperturbative terms, O($1/m_c^3$), in the expansion.

\section {About $\mu_\pi^2$}
The mesonic     
matrix element of the time dilation operator, $<D|\bar{Q} (i\vec{D})^2 Q|D> = 
\mu_\pi^2$ is not known accurately. An analysis based on sum rules yields
\cite{BALL}
\bb
\mu_\pi^2 = 0.5 \pm 0.1 \; {\rm{\;GeV}^2}.
\ee
This result is in agreement by a bound from
\cite{FERMI}, \cite{VOLIN}, \cite{KSV}, \cite{SUMRULES},
\bb
\mu_\pi^2 \ge \mu_G^2  \ge 0.37 \; {\rm{GeV}^2}
\ee
This inequality is of huge importance, and much controversy \cite{NEUBVIR}.
The derivation of this inequality using quantum mechanics, 
taught to me by Misha Voloshin in his 
excellent class on 
phenomenology, is simple. Consider the absolute positive quantity,
\bb
(\vec{\sigma} \cdot \vec{\pi})^2 = \vec{\pi}^2 - g \vec{\sigma} \cdot 
\vec{B}, \ee
then, obviously, the above inequality holds.
The field-theoretic inequality was later derived in \cite{SUMRULES}.
There, the sum rule for the pseudoscalar weak current, $J_5$ was written at
zero-recoil,
\bb
\frac{1}{2\pi}\int_0^\mu w^{(5)}(\epsilon)d\epsilon =
(\frac{1}{2m_c} - \frac{1}{2m_b})^2(\mu_\pi^2(\mu) - \mu_G^2(\mu)).
\ee
Since the structure function $w^{(5)}$ is non-negative, one arrives at the
conclusion that $\mu_\pi^2(\mu) \ge \mu_G^2(\mu)$.

Still, there are arguments in the literature that this inequality is broken by
the perturbative evolution of $\mu_\pi^2$. 
Consider figure  6.4 below
The figure represents the 
issue of concern regarding the 
impact of the perturbative evolution on the numerical value of $\mu_\pi^2$. 
I bring up the issue not only due to its  
possible numerical relevance, but also because it sheds light on the actual 
meaning of both $\mu_\pi^2$ and $m_Q$, and their relation to each other.
Recall that in the case of the operator $\bar{Q}Q$ (i.e. just the heavy quark 
mass diagram) we had a result (for the running of $m_Q$) which had no 
logarithmic dependance. In such a case, it is tempting to - due to the safe 
infrared limit - run $\mu$ to zero. For instance, this is what we {\it{have}} 
done in the case of $\bar{Q}Q$. As we have discussed above, however, this 
is an illegal procedure. As stated above, however, what is done in practice 
is the following: one takes known expressions, in our case one-loop QED 
corrections, and adds to these perturbative terms nonperturbative 
corrections expressed via certain matrix elements.  What is done in the 
case at hand is the following. 

\begin{figure} 
\centerline{\epsfysize=4cm\epsfbox{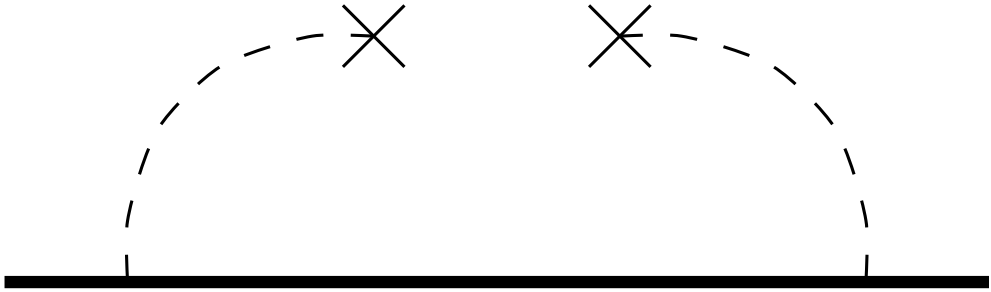}} 
\caption{The diagram corresponding to the perturbative evolution of 
$\mu_\pi^2$.} \end{figure}
From figure 6.4 it is easy to see that we can define the one-loop 
perturbative $\mu_\pi^2$ as just the infrared piece of the operator 
$\bar{Q}Q$, i.e.
\bb 
\bar{Q} \vec{\pi}^2 Q_{one-loop} = \frac{4 \alpha_s}{3 \pi} \mu^2 \bar{Q}Q
\ee
Now, as we know, this quantity should be subtracted from the actual 
$\mu_\pi^2$ (we cannot double count this piece!)
Numerically, one usually considers a value of $\mu$ relatively high so 
that $\alpha_s$ is small, but yet small enough so that the expansion in 
$\mu/m_c$ is good. To estimate the perturbative evolution, I will choose 
$\mu = 1 \; {\rm{GeV}}$. Doing so, the correction to $\mu_\pi^2$ is quite 
sizable - $0.15 \; {\rm{GeV^2}}$. The inequality stated above, $\mu_\pi^2 > 
\mu_G^2$ still is safe, however, as can be seen from eq. 6.27, at any scale.

%Another reason why the inequality holds is that the 
%operator $\mu_G^2$ also has a piece and its perturbative evolution 
%increases (according to its anomolaus dimension $\gamma_G =3$) towards 
%lower $\mu$! (Recall that we extracted $\mu_G^2$ from 
%the mass splittings of $B$ and $B^*$, and thus the normalization point here 
%is $\sim m_b \sim 4.82 \; {\rm{GeV}}$, 
%%thus the evolution of this operator must be 
%traced as well \cite{FGL}). The real runnning, at low $\mu$, however, is 
%given by the sum rule found in \cite{SUMRULES}, and the running of 
%$\mu_G^2$  is actually smooth in the infrared, and no contradiciton in 
%the inequality emerges.

For our purposes concerning the D lifetime, we realize, however, that the 
theoretical value of $\mu_\pi^2$ may not be 
completely crucial, for even if we take $\mu_\pi^2$ to have a very large 
perturbative evolution, so that $\mu_\pi^2 = 0.15 \; {\rm{GeV^2}}$, and so 
assuming the inequality held, $\mu_G^2 = 0.15 \; {\rm{GeV^2}}$ still we 
would have trouble fitting the lifetime, as can be seen in the next chapter.
\chapter{Putting it all together- The numerical prediction for $\Gamma(D)$ 
with leading order perturbative and nonperturbative corrections}
With the issues of numerics under our belts, we can finally check to see 
how our 
formula for $\Gamma_D$,
\bb
\Gamma(D) = \Gamma_0 (1 + A_1 \alpha_s - \frac{3 \mu_G^2}{2 m_c^2}
-\frac{\mu_\pi^2}{2 m_c^2}),
\ee
works.
We take the mindset that we push everything in the direction of 
agreement. This corresponds to a high $m_c = 1.4 \; {\rm{GeV}}$,  low 
$\mu_\pi^2 = 0.5 \; {\rm{GeV^2}} - 0.15 \; {\rm{GeV^2}} = 0.35 \; 
{\rm{GeV^2}}$
and low 
$\mu_G^2= 0.35 \; {\rm{GeV}^2}$ - I slightly lower the value of $\mu_G^2$ to 
reflect the fact that the inequality stated above still holds (due to the 
perturbative evolution mentioned in chapter 6).
Plugging everything in to the lifetime, then, we get:
\bb
\Gamma(D) = \Gamma_0 (1 - 0.24 - 0.27 - 0.09) = \Gamma_0 (0.40)
\ee
Recall that for $m_c = 1.4 \; {\rm{GeV}}$ 
we had perfect agreement 
with the lifetime 
{\it{without}} 
corrections, thus the 
corrections spoil agreement, lowering the lifetime about by half. 
To alleviate such a situation, we have to really push our numbers: 
one choice is to make $m_c = 1.6 \; {\rm{GeV}}$ $(1.6/1.4)^5\approx 2$). 
This raises the rate by about half, and solves our problem. 
However, doing so really kills any agreement with the $m_c$ extraction from 
the QCD sum rules. 
Under our philosophy, we wanted to take the sum rule determination as  
bedrock - especially, 
since as we will see there are exponentially small duality violation 
possibilities for 
sum rules, but not for decay widths. 

Another option is to try and drastically lower 
$\mu_\pi^2$. For instance, it might be that the 
extracted value of 
$\mu_\pi^2 = 0.35 \; {\rm{GeV}^2}$ then the perturbative evolution could 
knock 
the value down by say $0.15 \; {\rm{GeV}^2}$, and assuming that the 
inequality 
still holds, we still get a large theoretical deficit:
\bb
\Gamma(D) = \Gamma_0 (1.00 - 0.24 - 0.153 - 0.05) = \Gamma_0 (0.55)
\ee
And so it seems that even really pushing our nonperturbative corrections
to their lowest possible values, 
we 
are still in trouble.

\chapter{$1/m_c^3$ Corrections}
\begin{center}
{``How 
tragic is wisdom when it brings no profit to the wise''}
\end{center}

Now, it looks like we are stuck. How can we accomodate such a large 
width? A fairly non-imaginative option 
is that for some reason the - it is always an option 
- perturbative, or 
nonperturbative corrections are enhanced at the next order.
Thus, for instance, someone could try to go farther and calculate the 
$1/m_c^3$ 
corrections. This attempt was made in \cite{BDS}, and
in this section we 
review the calculation of these cubic terms.
As we saw in the 
last section in the numerical discussion, the hope is that the $1/m_c^3$ 
corrections can give us an 
increase in the width around 30\% to 50\%. Clearly, such a contribution will 
be due to an 
unsuspected enhancement in the coefficient of the $1/m_c^3$ operator. 

Since the total width is a 
Lorentz scalar, the only new operators relevant at the level of dimension 6 
are the four-fermion 
operators of the type:

\bb
O_6 = \bar{c} \Gamma q \bar{q} \Gamma c
\ee
where q stands for the light quark, and $\Gamma$ 
stands for Lorentz, and color structures.
There are two distinct 
sources of the $1/m_c^3$ corrections, just as there were for the $1/m_c^2$ 
corrections: operators of dimension 6 arising 
from the expansion of $\hat{T}$, and $1/m_c$ 
corrections in 
the D meson matrix 
elements of the operators $\bar{c} \sigma G c$ and $\bar{c}c$.

\section{The four-fermion operators at $O(\alpha_s)^0$ and in LLA}
A four-fermion 
operator appears in $\hat{T}$ in the zeroeth order in $\alpha_s$ in 
figure 8.1.
\begin{figure} 
\centerline{\epsfysize=4cm\epsfbox{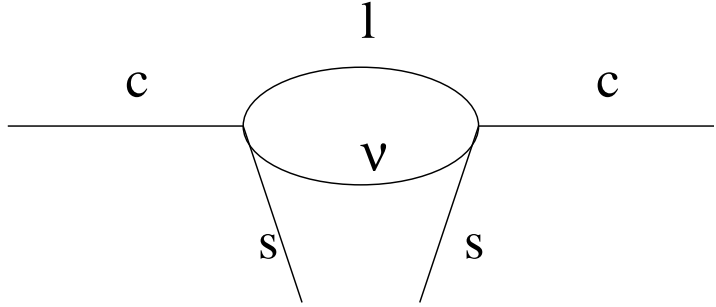}} 
\caption{The four-fermion term in the transition operator as it appears
at the level $\alpha_s^0$.} 
\end{figure}
%\begin{figure}
%\vspace{24cm}
%\special{psfile=FIG293_2.ps hscale=70 vscale=70
%hoffset=-10 voffset=-243 }
%\caption{
%$1/m^3$ graphs}
%\end{figure}
%\begin{figure}
%\vspace{24cm}
%\special{psfile=FIG293_1.ps hscale=70 vscale=70
%hoffset=-10 voffset=-243 }
%\caption{
%$1/m^3$ graphs}
%\end{figure}
The 
corresponding result can be read from Eq. $(17c)$ of ref \cite{SV},
\bb
{\rm{Im}} \; \hat{T}^{0} = -\frac{G_F^2 m_c^2}{8 \pi} 
[\bar{c_i}\Gamma_\mu c_k - (2/3) 
\bar{c_i} \gamma_\mu \gamma_5 c_k][\bar{s_k}\Gamma_{\mu} s_i],
\ee
here $\Gamma_\mu = \gamma_\mu (1-\gamma_5)$.
The expression above, however, yields zero for two reasons. First, in the 
factorization approximation, 
$\hat{T^0}$ corresponds to the annhilation contribution $c \bar s 
\rightarrow l \nu$, and 
our spectator here is not 
an s quark. Second, even if we deal with the $D_s$, the chiral structure 
yields a vanishing contribution when factorization is invoked, since under 
factorization $<D|\hat{T}^{0}|D> = 0$.
Thus this is not the contribution that we are looking for. But lets not be 
hasty.  The effective 
Lagrangian given above is at the virtuality scale, $\mu=m_c$.  The type of 
pre-asymptotic effect we 
are after is at a different scale, $\mu = \Lambda_{QCD}$. Under the 
rules of the OPE, we therefore 
must `evolve' the operator down to this infrared scale (`evolving' 
corresponds to sloshing some of the 
operator at scale $m_c$ into part of the coefficient at the scale 
$\Lambda_{QCD}$). Physically, this 
corresponds to including gluonic degrees of freedom, and thus, we can 
expect that the chirality, 
and hence vanishing character of the contribution which we obtained at 
scale $m_c$ will be 
different at the scale $\Lambda_{QCD}$. This contribution is calculated in 
ref. \cite{SV}, and is 

\begin{figure} 
\centerline{\epsfysize=4cm\epsfbox{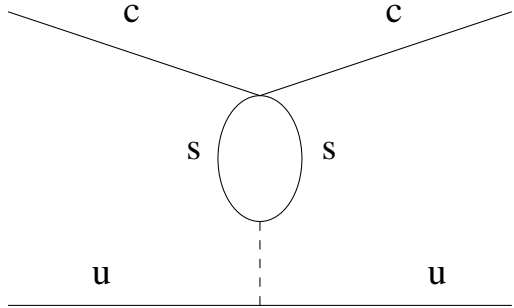}} 
\caption{The penguin graph for the four-fermion operator in $\hat{T}$.} 
\end{figure}

\bb
{\rm Im} \; \hat{T}^{1} = - \frac{G_F^2 m_c^2}{8 \pi} |V_{cs}|^2 [ - 
\frac{\alpha_s}{3 \pi} {\rm{ln}}
\frac{m_c^2}{\mu^2} (\frac{2}{3} \bar{c} \Gamma_\mu t^a c + \frac{1}{3} 
\bar{c} 
\tilde{\Gamma}_\mu t^a c) \sum_q \bar{q} \gamma_\mu t^a q],
\ee
here $\tilde{\Gamma}_\mu = \gamma_\mu (1+\gamma_5)$.

Notice that now, the light-quark current runs over {\it{all}} flavors and 
both left-handed {\it{AND}} right-handed fields appear, 
leading to the non-vanishing contribution of this 
result after factorization. 
This is a typical situation with the penguin contribution, and arises from 
the simple fact that the 
gluonic correction connects the $V\times A$ strange quark current to the 
pure vector background current. 
Above we have calculated the $1/m_c^3$ contribution arising from the 
logarithmic mixing of 
penguin operators and operators built from the light quark currents. 
Keeping only logarithms, however, may be an unjustified assumption, since 
$m_c/\mu$ is not large. Thus, in the next sections, we calculate the full 
$O(\alpha_s)$ contribution.

\section{Full $O(\alpha_s)$ calculation}
Here we divide the calculation of the full $O(\alpha_s)$ contribution 
into three seperate pieces.

\subsection{The `free' piece}
To start with, we calculate $1/m_Q^3$ pieces coming without any logarithms.
We get terms here in the 
same we got $1/m_c^2$ terms when we examined $\bar{Q} p^4 
\hat{p} Q$. And so, we can just start with the object:
\bb
{\rm{Im}} 
\; \hat{T_0} = \frac{G_F^2 |V_{cs}|^2 \bar{Q}(0) p^4 \hat{p} Q(0)}{384 
\pi^3}, 
\ee
where 
\bb
p_\mu = iD_{\mu} - g A_{\mu}.
\ee
There are two possible contributions coming from the first two 
terms of the expansion for $A_{\mu}$, recall that \bb
A_\mu = \frac{1}{2\times0!}x_\rho G_{\rho \mu} + 
\frac{1}{3 \times 1!} x_\rho x_\alpha (D_{\alpha} G_{\rho \mu})+...
\ee
(higher order terms will not contribute dimension 6 operators).
First, lets focus on the first\footnote{In ref.
\cite{BDS} this contribution was left out, but has minimal numerical 
effect anyway. The discrepancy between \cite{BDS} and \cite{GREMM} 
still exists even after this left out contribution is included}
term of 
the expansion of $A_\mu$, and so consider
\begin{eqnarray}
\bar{Q} \hat{p} p^4 Q &=& 
\bar{Q}(\hat{P}-\hat{A}) (P-A)^4Q \nonumber \\
 &=&\bar{Q}(\hat{P}-\frac{1}{2}\gamma_\mu x_\rho G_{\rho \mu})
(P-\frac{1}{2} x_\rho G_{\rho \mu})^4Q.  
\end{eqnarray}
As before, the strategy is to drag all $A_\mu$'s to the left, and 
$P_\mu$'s to the right. When $A_\mu$ stands to the left it gives zero 
(Q is taken at the origin, so since $A\sim x$, 
$AQ \rightarrow 0$), and when P stands to the right, 
the Dirac equation can be invoked.
Thus, we can already write
\bb
\bar{Q}(\hat{P})
(P-\frac{1}{2} x_\rho G_{\rho \mu})^4 Q.   
\ee
Continuing the procedure of pulling all gauge potentials to the left, we will
find ourselves in need of a few identities:
\bb
[P_\mu,A_\mu]=iD_\mu A_\mu = iD_\mu(\frac{1}{2}x_\rho G_{\rho\mu}) = 
\frac{i}{2} x_\rho D_\mu G_{\rho\mu},
\ee
\bb
[P_\nu,A_\mu]=iD_\nu A_\mu = \frac{i}{2}(G_{\nu\mu}+x_\rho(D_\nu 
G_{\rho\mu})), 
\ee
\bb
\gamma_\mu G_{\mu\nu}P_{\nu} = -\frac{1}{4}(\hat{P}\sigma G - \sigma G \hat{P})
.\ee
Using these identities, the final result for the first term of the gauge 
potential expansion is \bb
\bar{Q}\gamma_\rho D_\alpha G_{\alpha\rho}Q.
\ee

The second term in the expansion of the gauge potential also will give us
contributions $O(1/m_Q^3)$.
Again, consider
\begin{eqnarray}
\bar{Q} p^4 \hat{p} Q& = & \bar{Q} (P-A)^4 (\hat{P}-\hat{A}) Q\nonumber \\ 
& = & 
\bar{Q}(P^4 - P^2 P\cdot A - P^2 A\cdot P - 
A \cdot P P^2 - P\cdot A P^2)(\hat{P}-\hat{A})Q
\end{eqnarray}
We drop the term $\sim P^4\hat{P}$ since it gives no terms $O(1/m_Q^3)$, and
also drop the term $\sim A \cdot P P^2$ since $A$ here is completely to 
the left.
Thus, there are three terms which we need to examine which will give
$O(1/m_Q^3)$ contributions:

\begin{eqnarray}
\rm{Term} \; 1 &:& \bar{Q}(-P^4 \hat{A})Q \nonumber \\
\rm{Term} \; 2 &:& \bar{Q}(-P^2 [P_{\mu},A_{\mu}] \hat{P})Q \nonumber \\
\rm{Term} \; 3 &:& \bar{Q}(-2P^2 A \cdot P \hat{P})Q \nonumber \\.
\end{eqnarray}
To get the results of each term, one merely needs to know the few simple
commutators, easily worked out above in the Fock-Schwinger gauge.
The results are, term by term,
\begin{eqnarray}
\rm{Term} \; 1 &:& 
\frac{4}{3} \bar{Q}(D_\gamma G_{\gamma\mu} \gamma_\mu) Q m_Q^2
\nonumber \\
\rm{Term} \; 2 &:& 
-\frac{2}{3} \bar{Q}(D_\gamma G_{\gamma\mu} \gamma_\mu) Q m_Q^2
\nonumber \\              
\rm{Term} \; 3 &:& 
\frac{4}{3}  \bar{Q}(D_\gamma G_{\gamma\mu} \gamma_\mu) Q m_Q^2.
\end{eqnarray}             

Adding the results from both the first and second terms of
the expansion of the gauge
potential we get the transition operator:
\bb 
\hat{T}|_{free} = -i G_F^2 |V_{cs}|^2 \frac{\alpha_s}{32 \pi^2} m_c^2 \bar{c} 
\gamma_{\rho} 
t^a c \bar{q} \gamma_{\rho} t^a q,
\ee
where we have used the equation of motion,
\bb
D^{\alpha} G^{a}_{\alpha \mu} = -g \bar{q} \gamma_\mu t^a q.
\ee
This piece contributes, upon factorization (see later in this
section)
\bb
\frac{\Delta\Gamma}{\Gamma_0} = \frac{4 \alpha_s \pi f_D^2 M_D}{3 m_c^3}.
\ee

Now we calculate the full logarithmic contribution referred to above. 
It will carry with 
it the piece (infrared divergent) proportional to log 
that we treat with the OPE cutoff prescription, and will also carry a 
finite piece.

\subsection{The `Log' piece}
There 
is another contribution 
in the expansion of $\hat{T}$ arising from $O(DG)$ terms in the expansion of 
the light quark 
propagator. 
The origin of the term 
proportional to $D_\alpha G_{\alpha\mu}$ was made evident in section 4. 
Here we need to keep 
the mass of the s quark finite, however, since we will use it as an infrared 
regulator (we will 
perform the proper OPE, subtracting out from our coefficient the soft 
contribution form 0 to 
$\mu$).  Keeping the mass of the s-quark finite 
results in the introduction of McDonald 
functions into the propagator. 
The result is:
\begin{eqnarray}
S_q(x,0) & = & \frac{-i m^2}{4 \pi^2} 
\frac{K_1(m \sqrt{-x^2})}{\sqrt{-x^2}} 
\nonumber \\
& - & \frac{m^2 \hat{x}}{4 \pi^2 x^2} K_2(m\sqrt{-x^2}) +
\frac{\tilde{G_{\rho\lambda}}}
{8\pi^2} \frac{m K_1(m\sqrt{-x^2})}{\sqrt{-x^2}}(x_\rho \gamma
_\lambda \gamma_5)   \nonumber \\
& + & \frac{G_{\rho \lambda}}{16 \pi^2} m K_0(m\sqrt{-x^2})\sigma_{\rho\lambda}
\nonumber \\
& + & \frac{2}{3} g \frac{i}{32\pi^2} (2 K_0(m\sqrt{-x^2})D^\alpha G_{\alpha
\rho}\gamma_\rho - (D^\alpha G_{\alpha \rho} \hat{x} x^\rho 
\nonumber \\
& + & x^\gamma x^\alpha D_\gamma G_{\alpha \rho} \gamma^\rho - 3i x^\gamma
x^\alpha D_\gamma \tilde{G}_{\alpha \rho} \gamma_\rho \gamma_5) \frac{m K_1
(m\sqrt{-x^2})}{\sqrt{-x^2}}) + \cdot\cdot\cdot
\end{eqnarray}
Due to the introduction of the Mcdonalds functions, we will have integrals of
the type
\bb
\int [\frac{K_n(m\sqrt{(-x^2)})}{(-x^2)^p}] \; e^{ipx} \; d^4x
\ee
which can be found in ref. \cite{BB}.

\begin{figure} 
\centerline{\epsfysize=4cm\epsfbox{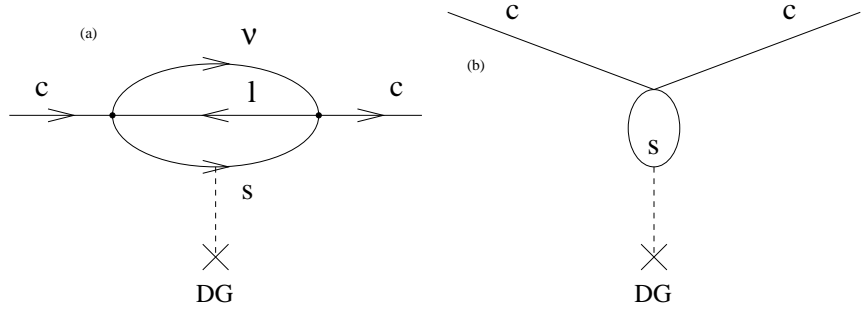}} 
\caption{The diagram with the $DG$ term in the s-quark line.} 
\end{figure}

To get the $O(1/m_Q^3)$ pieces, we isolate the piece of the propagator
proportional to $DG$ and one gamma matrice, see figure 8.2. The singular 
nature of the
contribution can be understood, since $K_n(m_s(\sqrt{-x^2})) \sim {\rm{ln}}
[m_s(\sqrt{-x^2})]$.

To avoid the infrared singularity, which shouldn't be included in the 
proper OPE calculation anyway, we calculate the graph of figure 8.3b, and 
subtract it from the graph of figure 8.3a. In 
this graph (10b), the lepton pair is hard - pointlike - and the s quark is 
soft. After subtracting the OPE infrared piece, we 
get the result, \begin{eqnarray}
\hat{T}_{\rm{ln}} &  = & i G_F^2 |V_{cs}^2| (\frac{\alpha_s}{72 \pi^2} m_c^2
({\rm{ln}}\frac{m_c^2}{\mu^2} + \frac{2}{3})) \nonumber \\ 
& + & (2 \bar{c} \Gamma_\mu t^ac + \bar{c} \tilde{\Gamma}_\mu t^a c) 
\sum\bar{q} \gamma_\mu t^a q
\end{eqnarray}
which gives a contribution aftor factorization,
\bb
\frac{\Delta \Gamma}{\Gamma} = - \frac{16 \pi \alpha_s}{9 m_c^3}
({\rm{ln}}\frac{m_c^2}{\mu^2} + \frac{2}{3}) f_D^2 M_D.
\ee

Lastly, we cannot forget to include pieces proportional to $DG$ coming 
from the expansion of 
$1/m^2$ operators. (Note, it is exactly how we picked up a piece of the 
$1/m^2$ contribution but 
there are no $1/m^3$ terms coming from $\bar{c}c$).

\subsection{The `$1/m^2$' piece}
Just as there were contributions of $1/m^2$ pieces coming from leading order 
terms  so also are there 
$1/m^3$ pieces originating from next to leading order terms.
Previously, 
we expressed 
$<D| \bar{c}\frac{i}{2} \sigma G c|D>$ in terms of the $D^* D$ mass splitting. This 
result is only valid to the leading order in $1/m_c$. Let us observe that the spin splitting yielding $M_{D^*}^2 - 
M_D^2$ is determined by the following terms in the heavy quark Hamiltonian:
\bb
\Delta H = \frac{1}{2 m_c} 
\vec{\sigma} \cdot \vec{B} + \frac{1}{4 m_c^2} \vec{\sigma} \cdot \vec{E}  
\times \vec{\pi},
\ee
where $\vec{B}$ and $\vec{E}$ are the 
chromomagnetic and chromoelectric fields respectively ($\vec{B} = g\vec{B}^a 
t^a$ and $\vec{E} = g\vec{E}^a t^a$.)
To leading order
\bb
\Delta_D = \frac{3}{4} (M_{D^*}^2 - M_D^2) = -<\vec{\sigma}\cdot\vec{B}> = 
0.405 \;  {\rm{GeV^2}}.
\ee
At the level of $1/m_c$ the second term in the heavy quark Hamiltonian becomes important in $\Delta_D$, as well 
as the second-order iteration in $(2 m_c)^{-1} <\vec{\sigma}\cdot \vec{B}>$.
Assuming that both effects are of the same order of magnitude, we can roughly estimate the matrix 
element $(2m_c^{-1})<\vec{\sigma}\cdot \vec{E} \times \vec{\pi}>$ as the difference between $\Delta_D$ and 
$\Delta_B$:
\bb
|(2 m_c)^{-1} <\vec{\sigma} \cdot\vec{E} \times \pi>| \le 
\Delta_D - \Delta_B \sim 0.04 \; {\rm{GeV}}.
\ee
Next observe that
\bb
\frac{i}{2} \bar{c} \sigma_{\mu\nu} G_{\mu\nu} c = - \bar{c} \vec{\sigma}
\cdot \vec{B} c - \frac{1}{m_c} \bar{c}
\vec{\sigma} \cdot \vec{E} \times \pi c - \frac{1}{2m_c}\bar{c}(D_i E_i) c.
\ee
The last term in the above equation reduces to the four-fermion operator which we take into account explicitly. The 
second term will be estimated as an uncertainty in the expression relating $<\frac{i}{2} \sigma G>$ to $\Delta_D$:
\bb
\mu_G^2 = <\frac{i}{2} \sigma G> = \Delta_D 
\pm 2 (\Delta_B - \Delta_D) - (2m_c)^{-1} 4 \pi \alpha_s <c 
\gamma_\mu t^a c \bar{q} \gamma_\mu t^a q>.
\ee
Using 
factorization 
for the $O_6$ term above, and the same values for the 
parameters as above, we get $+0.01$ for 
the contribution for $O_6$, so that
\bb
\mu_G^2 = \Delta_D \pm 2(\Delta_B - \Delta_D) = 0.42 \pm 0.08 \; 
{\rm{GeV^2}}.
\ee
As for 
$\mu_\pi^2$, it was shown in 
\cite{FGL} that the sign of the $1/m_c$ correction is negative. We will 
assume that the error bars in the numerical value of 
$\mu_\pi^2$ itself give the estimate of its $1/m_c$ contribution.

\section{Tallying the result}
In estimating the numerical value of the $1/m_c^3$ contribution we choose 
$f_D =170 \; {\rm{GeV}}$, $\alpha_s=.3$, and $m_c =1.4 {\rm{GeV}}$. 
Adding up all contributions, we get \bb 
\frac{\Delta\Gamma}{\Gamma} = -0.06 \pm 0.06 + 0.03
\ee
here the first number is due to the four-quark terms in the transition operator, the second is due to the uncertainty of $O_G$, and the third is due to $O_\pi$. Unfortunately, the result doesnÕt help our situation. Here we hardly have the 50 percent 
contribution we were looking for.

So it seems that again we are stuck - is there no way out?

\chapter {Ways out}
\begin{center}
{``Deny Everything'' - Hunter S. Thompson}
\end{center}
It appears that, 
in the case of the semileptonic decay of the D meson, all of our methods of 
attack have failed. What 
could have gone wrong? In this section, we review the case up to now,
focusing on possible solutions. Some possible solutions 
were reviewed in previous sections, and relied on increasing or 
decreasing relevant numerical parameters. There, we found it is clear 
that we 
cannot increase or decrease the numerical value of any parameters in the 
expression for $\Gamma(D)$ to obtain agreement with experiment - at least
not 
at the cost of creating even deeper puzzles in other charm phenomena. 
Also, it seems clear that the way out of this problem is not due to the 
perturbative series. In previous sections we saw quite clearly that the 
next order perturbative corrections make minimal impact. What other 
possibilities are left? 

One approximation that was made in chapter 8
was 
the factorization approximation in the evaluation of the dimension 6 
matrix elements. Corrections to the factorization approximation were 
reviewed in \cite{BLOKSHIF}. There, corrections were shown to be 
proportional to $1/N_c$ and seem to be small. Remember, we 
would need for the corrections to be quite large. 

The next possibility 
might be that dimension 7 operators could save the day. In principle, 
this may happen since the expansion parameter is $\approx 
\sqrt{\mu_G^2}/m_c \approx .5$ and is of order unity. However, since 
the correction due to $1/m_c^3$ terms is  of order 10 percent, this seems 
unlikely - not to mention, it would involve a very painstaking calculation! 

The most 
likely solution, however, does lie in the fact that our nonperturbative 
expansion parameter {\it{is}} large. This idea is tied conceptually
to the fact that the kinematics of the problem at hand are essentially 
Minkowskian. One justifies an OPE based procedure by keeping in mind an 
analytical continuation. In the problem of the semileptonic width this 
may be a continuation of the lepton pair - one considers the transition 
operator at such momenta where one is actually off the cuts 
corresponding to production of the hadronic states - in the Euclidean 
domain. The prediction on the cuts is made by invoking dispersion 
relations, in full analogy with what is usually done in the problem of 
the total $e^+ e^-$ annhilation. In general, one can analytically 
continue in some auxiliary momenta which has nothing to do with any of 
the physical momenta. 

Whatever analytic continuation is done, the prediction for each given 
term in the $1/m_c$ expansion refers to the Euclidean domain and is 
translated to the Minkowski domain only in the sense of averaging which 
occurs automatically through the dispersion relations. If the integrand 
is smooth, however, we can forget about the averaging because in this 
case smearing is not needed. This is what happens, in particular, with 
the total hadronic cross section in $e^+ e^-$ annhilation at high 
energies - quark-hadron duality sets in and the OPE-based consideration 
yields the value of the cross section at a given energy, locally (without 
smearing). At what energy release is the integrand smooth and can the 
terms in the $1/m_c$ expansion be predicted locally? The existing theory 
gives no answer to the question, but the problem at hand seems to suggest 
that the duality limit sets in well above $1.4 \; {\rm{GeV}}$ 

In the next section, I review the work done in \cite{CHIBISOV} regarding 
duality violations. This work was the first real attempt to try and 
capture the essence of QCD violations, and provides probably the best 
framework with which to escape from the puzzle at hand.

\chapter {Duality violations}

\begin{center}
{``Dave - you can't just learn this subject by reading papers about it, 
you {\it{have}} to solve some problems by hand!'' - Misha Shifman} 
\end{center}

\begin{center}
{``You don't get it? Sit down with the book - AND A PENCIL IN HAND! - and 
work out every intermediate step, and you'll get it. Good for the soul 
you know....'' - Serge Rudaz}
\end{center}

Depressing as our failure may seem at first, we should not look at this 
result as a failure, but instead as an opportunity. In fact, the D is an 
ideal testing ground for duality violations. 
Our QCD based calculations have failed to describe the hadronic dynamics,
lending evidence to the idea that there must be a contribution which 
we have 
left
out. In this section, I discuss the concept of duality, its violations, and its 
relation to heavy quark physics - in particular the problem of the 
semileptonic D decay.

The concept of duality was introduced long 
ago \cite{WEINBERG}. Essentially, its meaning is this: we calculate 
amplitudes with quarks and gluons, and 
compare our results to hadronic observables. If duality holds, then the 
complicated hadronic dynamics involved in a given 
process yield essentially little effect - the most important effects come just 
from the simple point-like interaction of the quarks and gluons. In 
accounting, say, for the effect of Fermi motion - 
including $\mu_\pi^2$, or in general any other operators, we take a stab at 
including 
some of these hadronization effects - from the quark-gluon side. Thus, 
each time we include some new operator, 
etc. into our calculation we redefine what we mean by duality. In the 
present problem of D decay,
our definition of duality consists of the expansion in $\alpha_s$ and 
$1/m_c$ corresponding to the hadronic lifetime. We 
prayed that this quark-gluon calculation was dual to the hadronic decay 
width. Well, it 
looks like its not! Thus, we need to stretch ourselves, and find some 
contribution which 
we have left out.  Clearly it is a difficult task. Since we have the first few 
terms in both 
the perturbative and nonperturbative expansions, the effect that we are 
looking for will 
have to come from somewhere deeper. 
The perturbative expansion goes like $\sim \alpha_s/\pi$ which is numerically
$\sim 0.1$. The nonperturbative series goes something like
$\sqrt{<\mu_G^2>/m_c^2} \sim 0.5$, so the nonperturbative series seems to be
most suspect if our duality analysis targets poor convergence properties of our
expansions.
The first suggestion of this type of 
contribution came 
in ref. \cite{SMILGAINST}.
\section{Exponential terms and asymptotic series}
Consider the series $\sum_n a_n n!$. This is an example of an asymptotic 
series: for a 
few first terms the series homes in on a limiting value, but then proceeds 
to skate away
with the inclusion of higher order terms. Of course, including all the terms 
(for the problem at hand) we get a finite 
result. In the perturbative and nonperturbative expansions we deal with 
this  
type of series. Actually, it is not a scary thing that we deal with 
asymptotic series: hopefully, it can be the case in some 
situations that the impossible task of calculating higher order terms in a 
given expansion 
is indeed a groundless task. Perhaps in a given case the optimal order is 
$n=2$ or 3, and we 
can forget about the higher order terms - the series should be truncated 
here. This 
might be just the case with the D meson - not a bad hunch, since we have 
experience leading us to believe $1/m_c$ is a large expansion parameter. 

Truncating the series at a finite 
order, we  introduce an exponential error. This can be seen 
in the following way. Suppose we have some function, $f(x)$ represented as the
expansion,
\bb
f(x) = a_0 + a_1 x + a_2 x^2 + ....
\ee
if the expansion is factorially growing, the coefficients can be written like
$a_n \sim c^n n!$ (in QCD, c is roughly $\mu$).
Now, as stated above, the series converges, but then starts to diverge again at
an optimal n. Roughly this n occurs when
\bb
x \sim (c^n n!)^{1/n}.
\ee
We want to take into account the size of the last term neglected. Using
Stirling's formula,
\bb 
n! \sim \sqrt{2 \pi n} \;  n^n e^{-n}, 
\ee
it is easy to see that the last term neglected is $\sim 
{\rm{exp}}(-x/c)$.
Thus, truncating the series in $\mu/m_Q$ we get an exponential accuracy 
$\sim {\rm{exp}}(-\rho m_Q)$ where $\rho \sim 1/\Lambda_{QCD}$ 
is some infrared distance.

\section{Physical picture of the exponential terms}
Surprisingly, there is a clear physical picture of the exponential 
error \cite{SHI}.
Consider figure 10.1.
\begin{figure} 
\centerline{\epsfysize=5cm\epsfbox{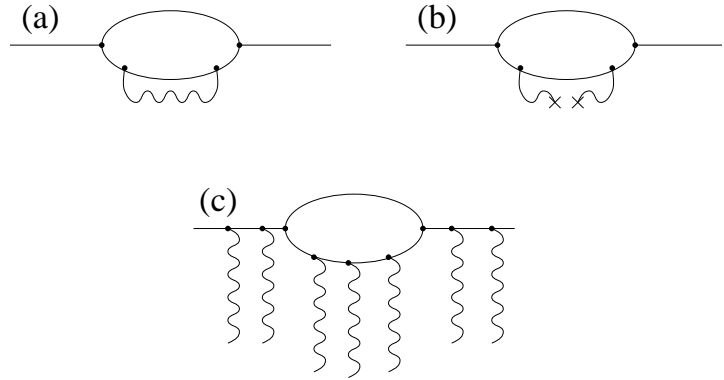}}
\label{BLA}
\caption{Possible diagrams contributing to heavy quark decay in a toy 
model.} 
\end{figure}
In discussing the figures, let me reemphasize some of the points I 
have made earlier  
regarding the OPE. First, consider figure 10.1a. Figure 10.1a is a usual 
perturbative diagram where the quark 
emits and reabsorbs a gluon. As is typical of perturbative graphs, a large 
energy release is carried by one or at most two quanta. Figure 10.1b is the 
nonperturbative `cut' diagram of figure 10.1a necessitated by 
the OPE. The cut line (cut since we have integrated down to our infrared 
point $\mu$) now 
becomes an operator, in particular, $\mu_\pi^2$.  Figure 10.1b. represents 
the diagrams which we 
normally take into account, and have, in the case of the semileptonic D 
lifetime, where one or at most two lines become soft. Figure 10.1c 
represents an 
additional contribution, the kind which gives us the exponential effect that 
we are looking for.
Here, we have  neither one or two hard lines nor one or two soft lines 
but rather, a 
hard contribution transmitted by ${\it many}$ lines, so that each line is soft. 
The situation is one where 
a hard momenta is shared by a coherent, possibly classic field 
configuration.
As we will see, by using instantons - just a solution to the classic equations 
of motion  - we will be 
able to model the exponential contribution which is conceptually related to 
figure 10.1c.  
Actually, these type of exponential terms were known long ago since the 
early days of QCD \cite{SVZ}. 
There, they were essentially disregarded as most of the early QCD 
application, e.g. sum 
rules, are of  Euclidean nature (dispersion relations are used).  In 
typical 
heavy quark cases we deal with Minkowski type kinematics (no 
dispersion 
relations, but instead a direct analytic continuation to the physical cut), 
and here the 
nature of the truncation error differs drastically - the exponentially 
decaying truncation error in the Euclidean domain
becomes oscillatory in the Minkowski domain, 
suppressed only by powers of our expansion parameter.

Our task then is simply to develop some methods to generate this type of 
exponential term. In fact, the exponential terms appear in the same way as the
power terms before them. A concrete example is given in \cite{POLEMASS}. There,
it is shown that the truncation of the factorially diverging series in 
$\alpha_s$ leads to an exponential error
\bb
\rm{exp}(-\frac{8\pi}{b \alpha_s(Q^2)}) \sim \frac{\Lambda_{QCD}^4}{Q^4}.
\ee
Thus, from the exponential error involved in truncation of the 
perturbative series, we can see the
presence of power terms. Here, we consider the series of power like terms. This
too is an asymptotic series \cite{SHI}, and its truncation will give an
exponential contribution of the type we are interested.
\section{Generating the exponential contribution}
To illustrate the main details of the instanton calculations, lets outline the 
practical motivation 
for the inclusion of the corresponding effects from the general perspective 
of the short distance 
expansion. Consider a generic two point function $\Pi(Q^2)$, the 
polarization operator of two 
vector currents (the Lorentz structure of the currents is irrelevant).
\bb
\Pi(Q) = \int d^4x e^{iqx} \Pi(x) = -\int d^4x e^{iqx} <G(x,0) G(0,x)>_0
\ee
where $G(x,y)$ is the quark Green's function in an external gauge field and 
averaging over the field 
configurations is implied - we work in Euclidean space.
The normal power type corrections occur when we consider the expansion 
of Green's functions 
near x=0 where the Green's function is singular.

We now examine the question `what happens when the propagator has a 
pole not near x=0?'. For 
example, consider the simplest finite-x singularity for the two-point 
function,
\bb 
\Pi(x) = \frac{1}{(x^2 + \rho^2)^\nu},
\ee
then,
\bb
\Pi(Q^2) = \int d^4x e^{iQx} \Pi(x) \sim e^{-Q \rho}.
\ee
These types of contributions are {\it{not}} seen in the normal OPE, and thus 
represent a violation of duality. Moreover, they mimick the type of 
behavior we expect from the higher order terms left out in the truncated 
series. 

Unfortunately, the physics of these duality violating terms brings us, again, 
to the main difficulty 
of QCD - we would like to know the background field fluctuations of the 
vacuum, but we don't. 
The problem enters since, although we do now the explicit form of the light 
quark propagator in 
the background of a distinct type of vacuum fluctuation - instantons (and this propagator has the form we want - 
poles off the origin), we 
don't know the density of instanton fluctuations in the infrared domain 
(i.e.  where  the poles are). Although, we know that  
instantons are not the dominant background fluctuation, we are not down on 
our 
luck completely, however, we will simply have to develop a model of the 
background fluctuations 
and fit it phenomenologically from known examples of possible duality 
violations.  
In the original paper \cite{CHIBISOV}, a two parameter model 
function was introduced, and 
fitted from two sources of duality violations: the semileptonic lifetime of 
the D, and the invariant hadronic mass distribution of $\tau$. The model 
was then used to explore possible duality violations elsewhere, e.g. B 
decays, $\alpha_s$ extraction from $\tau$ decays, and more. In the next 
sections, we review this model.

\section{Instantons and the OPE}
Before calculating the actual exponential terms in the real QCD case of  D 
decay, it will be useful to 
review certain universal aspects of the calculation in a toy model. The toy 
model generalizes the 
actual QCD calculation. 
Calculating the instanton contribution with the toy model, we clearly see 
that the instanton calculation can give three different types of 
contributions to the transition operator:

(i) Small-size instantons affect the coefficient functions of the OPE. We are 
not interested in these 
terms. They do not appear in our calculations below {\it{explicitly}} since 
our instanton density function 
excludes them. I will return to this issue when we consider the question of
the instanton density function in real QCD.

(ii) Terms proportional to $1/(m_Q\rho)$. They represent the instanton 
contributions to the matrix 
elements of various finite dimension operators that are present 
in the OPE. In principle, 
one could calculate each matrix element, pretending that the only 
contribution is from the one-instanton background. 
Such an attempt leads to results grossly violated in 
nature. 

(iii) The exponential terms. These terms, $\rm{exp}(-m_Q\rho)$ are what we 
are after.

Finding the exponential terms in the case of the heavy quark decay goes 
just like our 
example of the two-point function above, but with one subtlety - the 
external lines are colored 
objects, and thus feel the influence of the instanton. This {\it{could}} 
lead to considerable calculational 
difficulty. Fortunately there is a simplification. 
To see it, it is simplest to consider a toy model where all spins are 
neglected. The relevant features of the toy model 
translate to the case of real QCD.
\section{Instanton Calculation for a Toy Model}
The Lagrangian of the  toy model we consider has the form
\bb
L_W = h Q \bar{q} \phi + h.c.
\ee
which describes the decay of a heavy scalar quark Q into a 
massless quark q and a scalar photon $\phi$; the coupling 
h has dimensions of mass. Both quarks are in the spinor 
representation of the color group (SU(2) here). The basic 
strategy of the instanton calculation has been outlined 
above, here I work out details specific to the heavy quark case.

Consider the transition amplitude:
\bb
T = \frac{1}{2 M_D} <D |\hat{T}|D> = \frac{1}{2 M_D} 
<D|\int d^4x i T[L_W(x) L_W(0)]|D>,
\ee
see figure 10.2.

\begin{figure} 
\centerline{\epsfysize=3cm\epsfbox{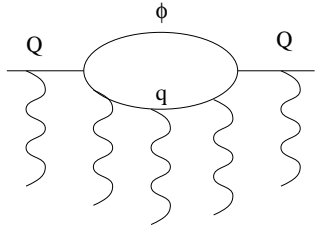}} 
\caption{The diagram representing heavy quark decay in a toy model in the 
instanton background field.} \label{F8}
\end{figure}

For our choice of the toy model Lagrangian,
\bb 
\hat{T} = i \int \bar{\tilde{Q}}(x) S(x,0) 
\tilde{Q}(0) G_{\phi}(x^2) e^{i m_c v x} d^4 x,
\ee
where $G_\phi$ is the propagator 
of the scalar photon, and 
$S(x,y)$ is the propagator of the massless scalar quark in 
the external (instanton) field.
The propagator of the massless scalar particle in the 
instanton background is known exactly:
\cite{BROWN}
\bb
S(x,y) = \frac{1}{4 \pi^2 (x-y)^2} 
(1 + \rho^2/x^2)^{-1/2} (1+\frac{\rho^2 (\tau^+ x)(\tau y)}{x^2 y^2})
(1+ \rho^2/y^2)^{-1/2},
\ee
where we have fixed the instanton center at $z=0$, and
\bb
\tau = (\vec{\tau},i); \; \tau^{+} = (\vec{\tau},-i); \;    
\tau_{\alpha}^{+} \tau_{\beta} = \delta_{\alpha\beta} + 
\eta_{\alpha\beta\gamma} \tau_{\gamma},
\ee
where $\vec{\tau}$ are the Pauli matrices acting in the color subgroup.

We now take a closer look at the final state quark propagator, rewriting it 
using the Feynman parametrization,
\begin{eqnarray}
S(x,y) & = & \frac{1}{\pi} \int^1_0 d\alpha 
[\alpha(1-\alpha)]^{-1/2} \frac{1}{\alpha (1-\alpha)(x-y)^2 + \rho^2 + 
\tilde{z}^2} \times \nonumber \\ 
 & + & \frac{1}{4\pi^2(x-y)^2}(1+ \frac{\rho^2 \tau^+ (x-z) \tau(y-z)}
{(x-z)^2(y-z)^2}) 
\sqrt{(x-z)^2} \sqrt{(y-z)^2}
\end{eqnarray}
where 
\bb
\tilde{z} = z - x \alpha - (1-\alpha)y.
\ee
In this form, the analytic structure of the propagator is clearer - we can now
easily
pick up the pole off the origin in our integrations, and avoid singularities at
the origin that give contributions which we are not interested in.
We want the pole:
\bb
x^2 = \frac{\rho^2 + \tilde{z}^2}{\alpha (1-\alpha)}.
\ee
After doing the integration over time,
at $m_c \rho \gg 1$, the remaining integrations are nearly Gaussian, 
and run over narrow intervals,
\bb
\vec{x}^2 \sim \frac{\rho}{m_c}; \; (\alpha - \frac{1}{2})^2 \sim \frac{1}
{m_c \rho}; \; 
(z - \frac{x}{2})^2 \sim \frac{\rho}{m_c}.
\ee
Thus, one performs 
the remaining 
integrations merely 
by evaluating all of the pre-exponential factors at the saddle point. In 
particular, consider the heavy quark external line. The heavy field 
$\tilde{Q}(x_0, \vec{x})$  can be written in the 
leading order as
\bb
\tilde{Q}(x_0,\vec{x}) = 
T e^{i \int^{x_0}_0 A_0(\tau,\vec{x})d\tau} 
\tilde{Q}(0,\vec{x}) + O(1/(m_c\rho))
= U(x) \tilde{Q}(0,\vec{x}) + O(1/(m_c\rho)),
\ee
where $U(x)$ is some 
complicated function (just the exponentiated, integrated instanton gauge 
potential). 
In principle, the complexity of 
the function $U(x)$ makes for an extremely difficult job of evaluating 
the transition operator. 
Fortunately, we are lucky, 
and in the saddle point approximation, $U(x)$ is just equal to 1! 
The heavy quark field enters at distances 
$\vec{x} \sim \sqrt{\rho/m_c} \ll \rho$ and, therefore the transition operator 
is finally proportional to $\bar{Q}(0) Q(0)$.
Collecting all remaining factors,

\bb
\hat{T}(k_0) = h^2 \bar{Q}(0) (\frac{G_\phi(-4\rho^2)}{2\pi^2\rho})
\int d\alpha d^3\vec{x} d^4z e^{-k_0 
\sqrt{(\rho^2+z^2)/(\alpha(1-\alpha) +\vec{x}^2}}Q(0)
\ee
where 
\bb
G_\phi(x^2) = \frac{1}{4\pi^2 x^2}
\ee
is the free scalar propagator. 
Performing the remaining Gaussian integrations, and taking the matrix 
element 
(note for scalar particles, 
$<H_Q|\bar{Q}{Q}|H_Q>=M_{H_Q}/m_Q$) we finally arrive at
\bb
T = -h^2 \frac{e^{(2im_Q\rho)}}{16 m_Q (m_Q^4)},
\ee
which gives us a contribution to the total width,
\bb
\Gamma_{scal}^I(\rho) = -h^2 \frac{\rm{sin}(2 m_Q \rho)}{8 m_Q^5 \rho^4}.
\ee
Note the oscillatory nature of the result upon continuation 
to Minkowski space. In applications, we set 
$\rm{sin}(m_Q\rho) =1$, 
since we use our model only as an approximate 
estimate of duality violations, and 
since, 
anyhow, the suspicion is that the strong dependance 
of $\rm{sin}(m_Q \rho)$ on the heavy quark mass, or position of the 
fixed-size instanton is somewhat artificial. The issue of the the remaining
$\rho$ dependance in the pre-exponential 
will be dealt with in the next section.

\section{The exponential contribution for $\Gamma_D$}
We now proceed to the actual calculation of the case of real D decay in the 
instanton model.
Let me outline the treatment.
First, we write the transition 
operator in the instanton background field. It has the same form as equation
(\ref{THAT}).
\bb
\hat{T} = \frac{G_F^2 |V_{cs}|^2}{2} \int \bar{\tilde{Q}(x)} 
\Gamma_\mu S((x,0),z,\rho) \Gamma_\nu 
\tilde{Q}(0) l_{\mu\nu}(x) e^{i(mvx)} d^4x \; d^4z \; d \rho \; D(\rho).
\ee
The Green function of the light quark, $S((x,0),z,\rho)$ is expanded in 
powers of $m_q$, and has 
the form:
\bb
S(x,y) = -\frac{1}{m_q}P_0(x,y) + G(x,y) + m_q \tilde{\Delta}(x,y) + 
O(m_q^2).
\ee

Technically, difficulties with the instanton model develops at the first step: 
the Green's function of 
the light quark in the background of one-instanton is not defined since the 
Dirac operator has a zero 
mode (zero modes are denoted above by $P_0$, and regulated by $m_q$). 
However, since the weak 
amplitude we consider contains left-handed quarks only, the problem of 
zero modes dissappears 
completely, albeit somewhat artificially. Hopefully, this not fully 
self-consistent procedure will work satisfactorily 
enough for our 
purposes. Of course, we have no right, in general, to believe that the 
one-instanton contribution 
will give us good phenomenology, still, since 
we scale our model from other duality violations, 
we can hope then that the duality violation hierarchy from decay to decay
is captured correctly.

Proceeding with the rest of the calculation  
see ref. \cite{CHIBISOV}
for an exhaustive discussion,
We arrive at the result:
\bb
\Gamma_{sl}^I = -\frac{2}{3} \Gamma_0 \frac{96 \pi}{(m_Q\rho)^8} \rm{sin}(2 
m_Q \rho) D(\rho).
\ee
Now, the real question of QCD dynamics enters in full. We need to integrate 
over the instanton 
size $\rho$, and an explicit representation of $D(\rho)$ is required. 
The calculation of $D(\rho)$ was first undertaken by t'Hooft
\cite{THOOFT}.
This instanton 
density function, however, is not what 
we want. In t'Hooft's calculation, only small-size 
instantons can be considered. His instanton density function took the form:
\bb
D({\rho}) = \rm{const.}(\rho \Lambda_{QCD})^b
\ee
where b is the first coefficient in the Gell-Mann Low function.
Integrating over this instanton density function leaves us not with the 
exponential quantity we are 
after, but instead power like contributions. 
This contribution was discussed above. 
Small size 
instantons are hard fluctuations. Taking them into account should in no 
way reflect the contribution 
of the series of higher order operators - yielding an exponential term. 
Indeed, considering the small-size instantons we are outside the validity of 
the standard heavy quark expansion (HQE). The standard 
HQE requires the decomposition of the heavy quark field in the form 
$Q(x) = {\rm{exp}}(im_Q v_\mu x_\mu) \tilde{Q}(x)$ 
which in the hard instanton background
becomes inapplicable, as well as the statement that heavy quark spin 
effects are 
suppressed by $1/m_Q$, and so on.

To get the exponential contribution coming from instantons
We need the density function 
in the infrared but, of course, it cannot be calculated there. We thus model 
$D(\rho)$ with the simplest idea we can think of:
a fixed size instanton:
\bb
D(\rho) = N \delta(\rho-\rho_0).
\ee
Actually, there is some reason to believe that this density function at least 
represents the character 
of the the soft instanton fluctuations. Results from the instanton liquid 
model 
\cite{SHURYAK}
show that the density 
function should have a sharp rise and steep fall-off. Thus, we can hope at 
least that our model correctly captures the essence of the real QCD 
instanton density function.
\section{Numerics of the instanton model}
After integrating over the instanton density function, we 
have the results of our duality model for the semileptonic decay
expressed in terms of two free parameters:
\bb
\Gamma_{sl}^I = - \Gamma_0 \frac{2}{3} N \frac{96\pi}{(m_c \rho_0)^8}
{\rm{sin}}(2 m_c \rho_0) = \Gamma_0 \frac{2}{3} N \frac{96 \pi}{(m_c 
\rho_0)^8}. 
\ee
Let me repeat that since the value of $\rm{sin}(2 m_c \rho_0)$ is sensitive to
how close the argument is to $n\pi$, a very model-dependant feature, we thus set
$\rm{sin}(2 m_c \rho_0) = 1$, and take the absolute value of the expresion,
taking the conservative point of view that our model at best determines an
{\it{uncertainty}} due to duality violations.

Now, we are at a crossroads. Unfortunately we cannot test whether the model we
have developed is capable of giving us a phenomenologically acceptable result.
The dependance of the result on two unfixed parameters requires us to extract
some knowledge of duality violations from some {\it{other}} processes. At 
this
stage of developement of both theory, and experiment, examples are rare. An
obvious place to go hunting is where hadronic processes have small energy
releases. In fact, there is one place - the invariant mass spectrum of 
hadronic $\tau$ decays - where we actually do see some preliminary signs of
duality violations. The issue is considered in detail in ref. \cite{CHIBISOV},
here I just sketch the results. 

\begin{figure} 
\centerline{\epsfysize=10cm\epsfbox{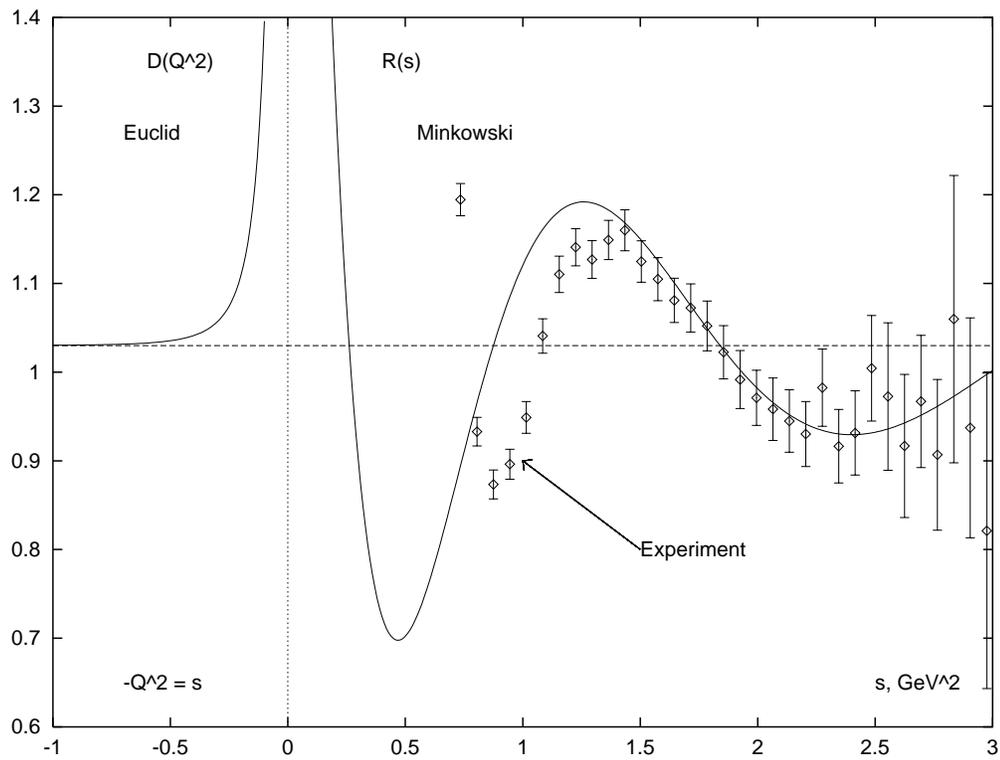}} 
\caption{Plot of $R^{(V-A)}(E)$ with Euclidean and Minkowski truncation
errors sketched on top of the data.} 
\label{F12}
\end{figure}

Figure 10.3 is the experimental plot of the quantity $R^{(V-A)}(E)$. From
the plot, it can be seen that there is, at the tail of the distribution
starting at about 2 GeV, evidence of some kind of oscillation $\sim 10 \%$. 
This oscillation
is exactly the behavior predicted by the Euclidean exponential error when
continued to the Minkowski domain. In ref. \cite{CHIBISOV}, this oscillation
was used as an input to determine one of the two free parameters of the
instanton density function. 

Because other sources of duality violations are scarce, we can fit the
remaining parameter of the model to the $\sim 50$\% duality violating 
contribution
required in the semileptonic D lifetime, and then use the model to
$\it{predict}$ violations in other processes. For example, calculating a
smattering of B lifetimes we end up with the result:
\bb
\frac{|\Gamma^I(b\rightarrow c\bar{c}s)|}{\Gamma_0(b\rightarrow c\bar{c}s)}
\sim 0.006
\ee
and 
\begin{eqnarray}
\frac{\Delta\Gamma^I}{\Gamma_0}(b\rightarrow c\bar{c}s) 
& \sim & 2\frac{\Delta\Gamma^I}{\Gamma_0}(b\rightarrow cud)
\sim 5\frac{\Delta\Gamma^I}{\Gamma_0}(b\rightarrow u\bar{u}d) \nonumber \\
& \sim & 16\frac{\Delta\Gamma^I}{\Gamma_0}(b\rightarrow cl\nu)
\sim 75\frac{\Delta\Gamma^I}{\Gamma_0}(b\rightarrow s\gamma) \nonumber \\
& \sim & 300\frac{\Delta\Gamma^I}{\Gamma_0}(b\rightarrow ulv)
,
\end{eqnarray}
thus, the deviations from duality in B decays are negligible.  

Let me briefly mention one other prediction the instanton 
model makes with, unlike in
the case of the B decays, more observable implications. If the approach above
is applied to the hadronic $\tau$ width, the deviations from duality are
estimated as
\bb
\frac{\Delta\Gamma^I(\tau\rightarrow 
{\rm{hadrons)}}}{\Gamma_0(\tau\rightarrow {\rm{hadrons}})} \sim 0.05
\ee
while this seems like a relatively small result, this $5\%$ uncertainty in the
width translates into a $\sim 30\%$ uncertainty in the value of
$\alpha_s(m_\tau)$. 
This is an interesting result in that the low energy
determinations of $\alpha_s$ currently disagree with measurements at the Z
peak - except for the case of $\alpha_s$ extraction from $\tau$ decays. 
Shifman 
\cite{SHIFALPHA} points out that this mismatch between low and high 
energy $\alpha_s$ extraction could be a prompt of new physics. 
The model outlined above should at the
very least make those who extract $\alpha_s$ from $\tau$ widths a bit 
nervous, and thus raise an eyebrow at the possibility of new physics 
being seen.
At the very least, here the strong case for large duality violations in $m_c$
where the energy release is $\sim 1.40 \; \rm{GeV}$ 
should make one suspect at least
minimal violations in $\tau$ where the energy release is $\sim 1.77 \; 
\rm{GeV}$.

Let me make one more closing remark concerning the semileptonic decay $b
\rightarrow c \tau \nu$. This decay was recently measured by \cite{ALEPH}. The
theoretical prediciton, including $1/m^2$ effects is found in \cite{FALK?}.
In the decay, there is an energy release of $\sim 1.6 \; \rm{GeV}$, thus, it is
tempting to consider that there is perhaps a duality violation on the order $5
- 10 \%$. Currently, the numerics on both the theoretical and expermental side
are not trustworthy enough to make an observation of duality violation, but it
very well may be a place to hunt for duality violations in the future, 
and thus possibly supply evidence for the utility of the instanton model.

\chapter{Conclusion}
\begin{center}
{``Physics is rich, but life is richer'' - Emil Akhmedov}
\end{center}
We have seen that the physics of the D meson decay is rich indeed! 
Calculation of the total decay 
width requires all of our theoretical might, and then some. OPE power 
corrections, normal 
perturbative terms, duality fluctuations - what a problem - and we are 
{\it{still}} evaded. 
In the end, we found that, at least at present, the semileptonic width of the D
meson is at best a theoretical kitchen for the study of duality violations.
This in itself is not a complete dissapointment. Many predictions from 
other
hadronic decays concerning CP violation, $\alpha_s$ extraction, etc. will come
in the future, and almost all, due to their Minkowskian nature, will come with
the built in assumption of duality. To wit, if we want to go from predictions
made by QCD without some flip assumption, and instead a well investigated one,
we need to study $D$ and $\tau$ decays to better understand the assumption of
duality. It is not ironic, but rather somewhat expected that in our duality
investigations, we here too come up against the problem of understanding QCD
dynamics in the infrared. Unfortunately, it seems that this is a problem which
will continue to defy solution for some time, and thus, it makes the job of
approaching QCD physics from the phenomenological side 
outlined in this thesis all the more important. It seems that in QCD, the
answer to Fermi's question is still yet to come.

\end{document}